\documentclass[letterpaper,twocolumn,prd,floats,preprintnumbers,showpacs]{revtex4}

\usepackage[english]{babel}
\usepackage[dvips]{graphicx}
\usepackage{amsmath,amssymb}
\usepackage{color}

\begin{document}

\newcommand{\etal}{et al.\ }

\newcommand{\cur}{{\chi}}
\newcommand{\abs}[1]{\vert#1\vert}
\newcommand{\bark}{{\bar{k}}}       
\newcommand{\fNL}{f_{\rm{NL}}}
\newcommand{\tauNL}{g_{\rm{NL}}}
\newcommand{\nadi}{n_{\rm{ad}}}
\newcommand{\nadiI}{n_{\rm{ad1}}}
\newcommand{\nadiII}{n_{\rm{ad2}}}
\newcommand{\niso}{n_{\rm{iso}}}
\newcommand{\ncor}{n_{\rm{cor}}}
\newcommand{\Omsd}{\Omega_{\cur  ,{\rm{dec}}}}
\newcommand{\mc}[1]{\mathcal{#1}}
\newcommand{\mr}[1]{\mathrm{#1}}
\newcommand{\sign}{{\mbox{sign}}}
\newcommand{\mb}[1]{{\mathbf{#1}}}
\newcommand{\nn}{\nonumber}
\newcommand{\sosc}{\cur_{\rm osc}}
\newcommand{\half}{\textstyle{\frac{1}{2}}}
\newcommand{\definit}{\stackrel{\mbox{\tiny{def}}}{=}}
\newcommand{\dNi}{{\frac{d N}{d p}}}
\newcommand{\dNii}{{\frac{d^2 N}{d p^2}}}
\newcommand{\dNiii}{{\frac{d^3 N}{d p^3}}}

\newcommand{\comment}[1]{ }
\newcommand{\jv}[1]{\textcolor{black}{{\bf JV: #1}}}

\title{Non-Gaussianity of the primordial perturbation in the curvaton model}
\author{Misao Sasaki}
\affiliation{Yukawa Institute for Theoretical Physics, Kyoto University, Kyoto 606-8503, Japan}

\author{Jussi V\"aliviita}
\email{jussi''DOT''valiviita''AT''port.ac.uk}
\affiliation{Institute of Cosmology and Gravitation, University of Portsmouth, Portsmouth PO1 2EG, UK}

\author{David Wands}
\affiliation{Institute of Cosmology and Gravitation, University of Portsmouth, Portsmouth PO1 2EG, UK}

\date{November 15, 2006}

\begin{abstract}
We use the $\delta N$-formalism to investigate the non-Gaussianity of
the primordial curvature perturbation in the curvaton scenario for the
origin of structure. 
We numerically calculate the full probability distribution function
allowing for the non-instantaneous decay of the curvaton and compare
this with analytic results derived in the sudden-decay approximation.
We also present results for the leading-order contribution to the
primordial bispectrum and trispectrum.
In the sudden-decay approximation we derive a fully non-linear
expression relating the primordial perturbation to the initial
curvaton perturbation. 
As an example of how non-Gaussianity provides additional constraints
on model parameters, we show how the primordial bispectrum on CMB
scales can be used to constrain variance on much smaller scales in the
curvaton field.
Our analytical and numerical results allow for multiple tests of 
primordial non-Gaussianity, and thus they can offer
consistency tests of the curvaton scenario.
\end{abstract}

\preprint{ YITP-06-33,\ \ ICG 06/35 }

\maketitle

%
%

\section{Introduction}

Most inflationary models give rise to a nearly Gaussian
distribution of the primordial curvature perturbation, $\zeta$.
Deviations from an exactly Gaussian distribution are
conventionally given in terms of a non-linearity parameter, $\fNL$
\cite{KomatsuSpergel}. The prediction for the non-linearity
parameter from {\it single-field} models of inflation is related
to the tilt of the power spectrum, $\fNL\sim n-1$ \cite{Maldacena}
which is constrained by observations to be much less than unity.
In principle, measurement of $\fNL$ would give a valuable test of
the inflation, but unfortunately such a tiny non-Gaussianity is
likely to remain unobservable. The current upper bound from the
WMAP three-year data \cite{Spergel:2006hy} is $\abs{\fNL} < 114$
while Planck is expected to bring this down to $\abs{\fNL}
\lesssim 5$ \cite{KomatsuSpergel} which is still orders of
magnitude larger than the prediction for single-field inflation.

On the other hand, {\it multi-field} models of inflation can lead
to an observable non-Gaussianity. One well-motivated example is
the curvaton model \cite{LWcurvaton,Enqvist:2001zp,Moroi:2001ct}:
In addition to the inflaton $\phi$ there would be another, weakly
coupled, light scalar field (e.g., MSSM flat direction
\cite{Enqvist:2002rf,Allahverdi:2006dr}), curvaton $\cur$, which
was completely subdominant during inflation. The potential could
be as simple as \cite{Bartolo:2002vf} $V=\half M^2\phi^2 + \half
m^2\cur^2$, where the energy density of $\phi$ drives inflation. At Hubble
exit during inflation both fields acquire some classical
perturbations that freeze in. However, the observed cosmic
microwave (CMB) and large-scale structure (LSS) perturbations can
result from the curvaton instead of the inflaton. If the inflaton
mass $M$ is much less than $10^{13}$~GeV then perturbations due to
the inflaton are much smaller than $10^{-5}$.

After the end of inflation the inflaton decays into relativistic
particles (``radiation''), the curvaton energy density still being
subdominant. At this stage the curvaton carries an isocurvature
(entropy) perturbation. The entropy perturbation between radiation
and curvaton is given by ${\cal S}_{\cur} \equiv 3(\zeta_\cur -
\zeta_r)$. Observations rule out purely isocurvature primordial
perturbations \cite{Enqvist:2000hp,Enqvist:2001fu}, but, so long
as the curvaton decays into radiation before primordial
nucleosynthesis, the entropy perturbation can be converted to an
adiabatic one.
This also requires that all the species are in thermal equilibrium and 
that the
baryon asymmetry is generated after the curvaton decays. If any of these
conditions are not met then the curvaton could also leave a residual
isocurvature perturbation \cite{Lyth:2002my},
but in practice the amplitude of isocurvature modes are severely
constrained by current data \cite{Trotta:2006ww,Lewis:2006ma,Bean:2006qz}
and for simplicity we will not consider this possibility
in this paper.

As the Hubble rate, $H$, decreases with time after inflation,
eventually $H^2 \lesssim m^2$, and the curvaton starts to oscillate
about the minimum of its potential. Then it behaves like pressureless
dust (with density inversely proportional to volume, $\rho_\cur
\propto a^{-3}$) so that its relative energy density grows with
respect to radiation ($\rho_r \propto a^{-4}$). Finally, the curvaton
decays into ultra-relativistic particles leading to the standard
radiation dominated adiabatic primordial perturbations \footnote{%
Had we not assumed negligible inflaton curvature perturbation,
$\zeta_\phi \approx 0$, some ``residual'' isocurvature would have
resulted if the curvaton was sub-dominant during its decay. This
would have led to an interesting mixture of correlated adiabatic
and isocurvature perturbations which was studied in
\cite{Ferrer:2004nv}. Following the guidelines of
\cite{Ferrer:2004nv} our calculations should be straightforward to
generalise. It should be noted that {\it observations do not rule
out} a correlated isocurvature component if it is less than 20\%
of the total primordial perturbation
amplitude \cite{Kurki-Suonio:2004mn}.%
}. However, this curvaton mechanism may, from the initially
Gaussian curvaton field perturbation, $\delta\cur_\ast$, create a strongly
non-Gaussian primordial curvature perturbation, $\zeta$. 
The non-Gaussianity is large if the energy density of the curvaton is
sub-dominant when curvaton decays. 
Since the amplitude of the resulting perturbation depends on the model
parameters (such as the curvaton mass $m$ and decay rate $\Gamma$),
the observational bounds on non-Gaussianity provide important
constraints on model parameters.

The objective of this paper is to calculate the probability density
function (pdf) of the primordial curvature perturbation in the
curvaton model. Since, in the early universe, all today's observable
scales are super-Hubble scales after inflation, we take advantage of
the {\it separate universe} assumption \cite{ST,WMLL} throughout the
calculations and employ the so-called $\delta N$-formalism
\cite{Starobinsky,SaSt,LR}. This allows us to determine the pdf fully
non-linearly (not just up to second or third order in the initial
field perturbations) so that it will carry all the information about
non-Gaussianity.

Generally we can expand any field
 \begin{equation}
 \varphi = \bar{\varphi} + \sum_{n=1}^\infty \frac{1}{n!} \delta_n \varphi \,.
 \end{equation}
We take the background field to be spatially homogeneous,
$\bar\varphi(t)$, and we will further assume that the first-order
perturbation, $\delta_1 \varphi(t,x)$ is a Gaussian random field,
consistent with what we expect from the linear evolution of initial
vacuum fluctuations. Thus the higher-order perturbations,
$\delta_n\varphi$ for $n>1$, will describe non-Gaussian
perturbations of any field.


The primordial perturbation can be described in
terms of the non-linear curvature perturbation on
uniform-density hypersurfaces \cite{LMS}
\begin{equation}
 \label{zetanl}
\zeta (t,{\bf x}) = \delta N (t,{\bf x}) +
 \frac13 \int_{\bar\rho(t)}^{\rho(t,{\bf x})}
 \frac{d\tilde\rho}{\tilde\rho+\tilde{P}} \,,
\end{equation}
where $\delta N$ is the perturbed expansion, $\tilde\rho$ the local density and
$\tilde{P}$ the local pressure.
We expand the curvature perturbation as
\begin{equation}
\zeta(t,{\bf x}) = \zeta_1(t, {\bf x})
 + \sum_{n=2}^{\infty}\frac{1}{n!}\zeta_n(t, {\bf x})\,,
\end{equation}
where the pdf of $\zeta_1$ is Gaussian as it is directly proportional
to the initial Gaussian field perturbation, but the higher-order
terms give rise to a non-Gaussian pdf of the full $\zeta$.
The non-linearity parameters $\fNL$ and $\tauNL$ are defined by
\begin{equation}
 \label{fNLtauNL}
\zeta = \zeta_1 + \frac35 \fNL \zeta_1^2 + \frac{9}{25}
 \tauNL\zeta_1^3 + {\cal O}(\zeta_1^4) \,,
\end{equation}
or, equivalently,
\begin{eqnarray}
\zeta_2 & = & \frac65 \fNL \zeta_1^2 \,,\label{fNLzeta}\\
\zeta_3 & = & \frac{54}{25}\tauNL \zeta_1^3 \,\label{tauNLzeta}\,.
\end{eqnarray}
The numerical factors $6/5$ and $54/25$ arise because in linear theory
the primordial curvature perturbation $\zeta$ is related to the
Bardeen potential on large scales (in the matter-dominated era,
md), $\Phi_{H \rm md}=(3/5)\zeta_1$, which implies
\cite{KomatsuSpergel,Okamoto:2002ik,Kogo:2006kh}
\begin{equation}
\frac35 \zeta
 = \Phi_{H \rm md} + \fNL \Phi_{H \rm md}^2 + \tauNL \Phi_{H \rm md}^3\,.
\end{equation}
We are specifically interested in non-linear quantities and, as it is
$\zeta$ not $\Phi_H$ that is non-linearly conserved for adiabatic
perturbation on large scales
\cite{Lyth:2003im,Rigopoulos:2003ak,Malik:2003mv,LMS,Langlois:2005qp},
we will take Eqs.~(\ref{fNLzeta}) and (\ref{tauNLzeta}) as our
fundamental definition of the primordial parameters $\fNL$ and
$\tauNL$, respectively.

If we write the primordial power spectrum as
\begin{equation}
\langle \zeta({\bf k_1}) \zeta({\bf k_2}) \rangle = (2\pi)^3 P(k_1)
\delta^3({\bf k_1}+{\bf k_2}) \,,
\end{equation}
then the leading order contributions to the bispectrum and
(connected part of the) trispectrum are given by
\begin{eqnarray}
\langle \zeta(&& \hspace{-6.5mm}{\bf k_1})\zeta({\bf k_2}) \zeta({\bf k_3})
\rangle
 \nonumber
\\
& = & \!\! (2\pi)^3 B({\bf k_1},{\bf k_2}) \delta^3({\bf k_1}+{\bf
k_2}+{\bf k_3}) \,,\\
\langle \zeta(&& \hspace{-6.5mm}{\bf k_1})\zeta({\bf k_2}) \zeta({\bf k_3})
 \zeta({\bf k_4})\rangle\nonumber
\\
& = & \!\! (2\pi)^3 T({\bf k_1},{\bf k_2},{\bf k_3}) \delta^3({\bf k_1}+{\bf
k_2}+{\bf k_3}+{\bf k_4}),
\end{eqnarray}
where
\begin{eqnarray}
B({\bf k_1},{\bf k_2}) \!\! & = & \!\! (6/5)\fNL \left[ P(k_1) P(k_2) + 2\,{\rm perms} \right]\!,\\
\!\!\!\!\!T({\bf k_1},{\bf k_2},{\bf k_3}) \!\! & = & \!\!
\nonumber\\
 (18/25)\fNL^2 && \hspace{-6.5mm}\left[ P(k_1)P(k_2)P(|{\bf k_1}+{\bf
k_2}|) + 23\,{\rm perms} \right] \nonumber\\
 + (54/25) \tauNL && \hspace{-6.5mm}\left[ P(k_1)P(k_2)P(k_3) + 3\,{\rm perms} \right]
\,.
\label{Tvsfnlandgnl}
\end{eqnarray}
The first term appearing in Eq.~(\ref{Tvsfnlandgnl}),
which gives the dependence of the
trispectrum on second-order perturbations, and hence $\fNL$, was given in
\cite{Boubekeur:2005fj}.
But there is also a term dependent upon the third-order
perturbation, and hence $\tauNL$, which appears at the same order, and has a
different dependence upon the four wavevectors.

Previous estimates of non-Gaussianity in the curvaton scenario have
been based on expansions up to second order in the curvature
perturbation.
We will go beyond previous analyses and calculate the contribution
of the third order term in Eq.~(\ref{fNLtauNL}) to the trispectrum.
We compare our analytic expressions for $\fNL$ and $\tauNL$ in the
{\it sudden-decay} approximation \cite{Bartolo,LR} with numerical
results where we include the gradual decay of the curvaton,
transferring energy from the curvaton to the radiation. Indeed using
our numerical code we are able for the first time to give the full
probability distribution for the primordial curvature perturbation
in both the sudden-decay and the {\it non-instantaneous decay} case.
We will calculate the skewness (third moment of the pdf) and
kurtosis (fourth moment of the pdf) as well as higher moments of the
fully non-linear probability distribution function.

This paper is organised as follows. In Sec.~\ref{sectcurvaton} we
relate the curvaton curvature perturbation $\zeta_\cur$ to the
initial field perturbation $\delta_1\cur$ at the beginning of the
curvaton oscillation. Then, in Sec.~ \ref{sectsudden}, we derive
in the sudden-decay approximation a non-linear equation that relates
the primordial curvature perturbation $\zeta$ (when curvaton has
decayed) to the curvaton curvature perturbation at the beginning of
curvaton oscillation, $\zeta_{\cur}$, (or to the Gaussian curvaton
field perturbation $\delta\cur_\ast$ at horizon exit). We write down
the full solution of this equation in the Appendix A. In Sec.~\ref{sectsudden}
we continue solving this equation order by order,
and deriving the non-linearity parameters $\fNL$ and $\tauNL$ in the
sudden-decay approximation. In Sec.~\ref{sectnumerical} we
describe our fully non-linear numerical approach.
We compare the numerical non-instantaneous decay results
for  $\fNL$ and $\tauNL$ to the sudden-decay approximation.
In Sec.~\ref{sectpdf} we
calculate the pdf of $\zeta$ both in the sudden-decay approximation
and in the non-instantaneous decay case. We show the skewness and
kurtosis of the pdf as a function of curvaton model parameters.
In Sec.~\ref{sectvariance} we add one possible complication
to the analysis. Namely, the variance of curvaton field perturbations
$\langle (\delta\cur)^2 / \bar\cur^2 \rangle$ on smaller
than observable scales is not directly constrained. However,
the large variance leads to large non-Gaussianity
(see e.g. \cite{LM97,LM06}) so that
the observational bounds on non-Gaussianity set an upper
limit to this small-scale variance. We derive
a quantitative equation that relates $\fNL$ and the
variance, and use this equation with WMAP third year
bounds. Finally, in Sec.~\ref{sectconclusions}
we present concluding remarks.

\section{Non-linearity of the curvaton perturbation}
\label{sectcurvaton}

When the curvaton starts to oscillate about the minimum of its
potential, but before it decays, the non-linear curvature perturbation
on uniform-curvaton density hypersurfaces is given by \cite{LMS}
\begin{equation}
 \label{zetacurv}
\zeta_\cur (t,{\bf x}) = \delta N (t,{\bf x}) +
 \int_{\bar\rho_\cur(t)}^{\rho_\cur(t,{\bf x})}
 \frac{d\tilde\rho_\cur}{3\tilde\rho_\cur} \,.
\end{equation}
Hence, the curvaton density on spatially-flat hypersurfaces is
\begin{equation}
\label{zetasigma}
 \left. \rho_\cur\right|_{\delta N=0} = e^{3\zeta_\cur} \bar\rho_\cur  \,.
\end{equation}
Assuming the curvaton potential is described by a quadratic
potential about its minimum, the energy density is given in terms of
the amplitude of the curvaton field oscillations
\begin{equation}
\rho_\cur = \frac12 m^2 \cur^2 \,.
\end{equation}

We expect the quantum fluctuations in a weakly coupled field such
as the curvaton at Hubble exit during inflation, $\delta\cur_*$, to be well
described by a Gaussian random field
(see e.g. \cite{Seery:2005gb,Lyth:2005qj,Seery:2006vu}). 
Hence we will write
\begin{equation}
\cur_* = \bar\cur_* + \delta_1 \cur_* \,,
\end{equation}
with no higher-order, non-Gaussian terms.

Non-linear evolution on large scales is possible if the curvaton
potential deviates from a purely quadratic potential away from its
minimum \cite{LythNG,nurmi}.  Thus, in general, the initial amplitude
of curvaton oscillations, $\cur$, is some function of the field value
at the Hubble exit; $\cur=g(\cur_\ast)$. (The curvaton potential is in
any case virtually quadratic sufficiently close to the minimum.)
%
%
Thus we have during the curvaton oscillation
 \begin{eqnarray}
 \bar\rho_\cur & = & \half m^2 \bar g^2 \,,\label{Eqbarrhocur}\\
\rho_\cur & = &  \half m^2 \left[\bar g + \sum_{n=1}^{\infty} \frac{1}{n!}g^{(n)}
\left(\frac{\bar g}{g'}\frac{\delta_1\cur}{\bar\chi}\right)^n\right]^2 \,,\label{Eqdeltarhocur}
\end{eqnarray}
where we used the relation $\delta_1\cur = g'\delta_1\cur_\ast$ and
wrote $\bar g \definit g(\bar\cur_\ast)$ and
$g^{(n)} \definit \left. \partial^{n} g(\cur) / \partial \cur^n \right|_{\cur = \bar\cur_\ast}$.

Substituting (\ref{Eqdeltarhocur}) and (\ref{Eqbarrhocur}) into
(\ref{zetasigma}) we obtain
\begin{eqnarray}
\label{Eqfullzetacur}
e^{3\zeta_\cur}
& = & \frac{1}{\bar g^2}\left[\bar g + \sum_{n=1}^{\infty} \frac{1}{n!}g^{(n)}
\left(\frac{\bar g}{g'}\frac{\delta_1\cur}{\bar\chi}\right)^n\right]^2 \,.
\end{eqnarray}

Order by order, we have from (\ref{Eqdeltarhocur})
 \begin{eqnarray}
 \delta_1\rho_\cur &=& m^2 g\delta_1\cur \,, \\
 \delta_2\rho_\cur &=& m^2 \left( 1 + \frac{gg''}{g^{\prime2}} \right)
 \left( \delta_1\cur \right)^2 \,,\\
 \delta_3\rho_\cur &=& m^2 \left( 3\frac{g''}{g^{\prime2}} + \frac{g
     g'''}{g^{\prime3}}\right)\left(\delta_1\cur\right)^3 \,,
\end{eqnarray}
and from (\ref{Eqfullzetacur})
\begin{eqnarray}
\label{zetas1}
 \zeta_{\cur1}
 &=&
\frac23 \frac{\delta_1\cur}{\bar\cur} \,,\\
 \label{zetas2}
 \zeta_{\cur2}
 &=&
 - \frac32 \left( 1 - \frac{gg''}{g^{\prime2}} \right)
 \zeta_{\cur1}^2 \\
 \label{zetas3}
 \zeta_{\cur3} &=& \frac92 \left( 1 - \frac32\frac{gg''}{g^{\prime2}}
  + \frac12 \frac{g^2g'''}{g^{\prime3}} \right) \zeta_{\cur1}^3
 \,.
\end{eqnarray}
Here and in what follows, we omit the bar from $\bar g$ and simply denote it
by $g$.

Using (\ref{zetas1}) and (\ref{zetas2}), we can express the
second-order skewness in terms of the effective non-linearity
parameter for the curvaton perturbation, analogous to
Eq.~(\ref{fNLzeta}),
\begin{equation}
 \label{fNLs}
 \fNL^{\cur} = -\frac{5}{4} \left( 1 - \frac{gg''}{g^{\prime2}} \right)
 \,.
\end{equation}
Hence we find $\fNL^\cur=-5/4$ for the curvaton $\zeta_\cur$ in the
absence of any non-linear evolution ($g''=0$). If the curvaton comes
to dominate the total energy density in the universe before it decays,
so that $\zeta=\zeta_\cur$, then this is the generic prediction for
the primordial $\fNL$ in the curvaton model, as emphasised by
\cite{LR}.

From the third-order term (\ref{zetas3}) we obtain a contribution to
the trispectrum of the curvaton perturbation, described by a
non-linearity parameter analogous to Eq.~(\ref{tauNLzeta}),
\begin{equation}
 \label{f2s}
\tauNL^{\cur} = \frac{25}{12} \left( 1 - \frac32\frac{gg''}{g^{\prime2}}
  + \frac12 \frac{g^2g'''}{g^{\prime3}} \right) \,.
\end{equation}

\section{Sudden-decay approximation}
\label{sectsudden}

Most analytic expressions for the primordial density perturbation in
the curvaton scenario assume the instantaneous decay of the curvaton
particles. In this section we will derive
an equation
for the non-linear curvature perturbation,
and then use it to find
the non-linearity parameters $\fNL$ and $\tauNL$ in this {\em
sudden-decay} approximation.

In the absence of interactions, fluids with a barotropic equation of
state, such as radiation ($P_r=\rho_r/3$) or the
non-relativistic curvaton ($P_\cur=0$), have a conserved curvature
perturbation \cite{LMS}
\begin{equation}
 \label{zetai}
\zeta_i
 = \delta N +
 \frac13 \int_{\bar\rho_i}^{\rho_i}
 \frac{d\tilde\rho_i}{\tilde\rho_i+P_i(\tilde\rho_i)} \,. 
\end{equation}

We assume that the curvaton decays on a uniform-total density
hypersurface corresponding to $H=\Gamma$, i.e., when the local
Hubble rate equals the decay rate for the curvaton (assumed
constant). Thus on this hypersurface we have
\begin{equation}
 \label{barrho}
\rho_r(t_{\rm dec},{\bf x}) + \rho_\cur(t_{\rm dec},{\bf x}) =
\bar\rho(t_{\rm dec}) \,, 
\end{equation}
where we use a bar to denote the homogeneous, unperturbed quantity.
Note that from Eq.~(\ref{zetanl}) we have $\zeta=\delta N$ on the
decay surface, and we can interpret $\zeta$ as the perturbed
expansion, or ``$\delta N$''. Assuming all the curvaton decay
products are relativistic, we have that $\zeta$ is conserved after
the curvaton decay since the total pressure is simply $P=\rho/3$.

By contrast the local curvaton and radiation densities on this decay
surface may be inhomogeneous and we have from Eq.~(\ref{zetai})
\begin{eqnarray}
\zeta_r &=& \zeta + \frac14 \ln \left( \frac{\rho_r}{\bar\rho_r}
\right) \,,\\
\zeta_\cur &=& \zeta + \frac13 \ln \left( \frac{\rho_\cur}{\bar\rho_\cur}
\right) \,,
\end{eqnarray}
or, equivalently,
\begin{eqnarray}
\rho_r &=& \bar\rho_r e^{4(\zeta_r-\zeta)} \,,\\
\rho_\cur &=& \bar\rho_\cur e^{3(\zeta_\cur-\zeta)} \,.
\end{eqnarray}

Requiring that the total density is uniform on the decay surface,
Eq.~(\ref{barrho}), then gives the relation
\begin{equation}
 \label{nlzeta}
(1-\Omsd) e^{4(\zeta_r-\zeta)} + \Omsd e^{3(\zeta_\cur-\zeta)} = 1 \,,
\end{equation}
where $\Omsd=\bar\rho_\cur/(\bar\rho_r+\bar\rho_\cur)$ is the
dimensionless density parameter for the curvaton at the decay
time. This simple equation is one of the main results of this
paper. It gives a fully non-linear relation between the primordial
curvature perturbation, $\zeta$, which remains constant on large
scales in the radiation-dominated era after the curvaton decays,
and the curvaton perturbation, $\zeta_\cur$ described in
Sec.~\ref{sectcurvaton}, in the sudden-decay approximation.
In the limiting case where $\Omsd\to1$ (i.e., the energy density of
the curvaton comes to dominate before it decays) we have
$\zeta\to\zeta_\cur$, but in general Eq.~(\ref{nlzeta}) gives a
non-linear relation between $\zeta$ and $\zeta_\cur$.

For simplicity we will restrict the following analysis to the
simplest curvaton scenario in which the curvature perturbation in
the radiation fluid before the curvaton decays is negligible,
i.e., $\zeta_r=0$. After the curvaton decays the universe is
dominated by radiation, with equation of state $P=\rho/3$, and
hence the curvature perturbation, $\zeta$, is non-linearly
conserved on large scales. With $\zeta_r=0$ Eq.~(\ref{nlzeta})
reads
\begin{equation}
 \label{nlzetasimpl}
e^{4\zeta} - \left[ \Omsd e^{3\zeta_\cur} \right] e^{\zeta} + \left[ \Omsd -1 \right] = 0 \,,
\end{equation}
which is a fourth degree equation for $X=e^\zeta$. In the Appendix A we
give the solution of this equation. Since we already know
$e^{3\zeta_\cur}$ as a function of the initial field perturbation
$\delta\cur_\ast$, we have now found a full non-linear
mapping of the Gaussian perturbation $\delta\cur_\ast$
to the primordial (non-Gaussian) curvature perturbation $\zeta$.
We can Taylor expand the solution (\ref{zetafullAppendix}) to find first, second,
and third order expressions or we can (re-)solve Eq.~(\ref{nlzetasimpl})
order by order as we do in the following subsections.

\subsection{First order}

At first order Eq.~(\ref{nlzeta}) gives
\begin{equation}
4(1-\Omsd) \zeta_1 = 3 \Omsd (\zeta_{\cur1}-\zeta_1) \,,
\end{equation}
and hence we can write
\begin{equation}
 \label{defr}
\zeta_1 = r \zeta_{\cur1}
 \,,
\end{equation}
where \cite{LWcurvaton}
\begin{equation}
 \label{defrSD}
r = \frac{3\Omsd}{4-\Omsd}
= \left.\frac{3\bar\rho_\cur}{3\bar\rho_\cur + 4\bar\rho_r}\right|_{t_{\rm dec}} \,.
\end{equation}

\subsection{Second order}

At second order Eq.~(\ref{nlzeta}) gives
\begin{multline}
4(1-\Omsd) \zeta_2 - 16(1-\Omsd) \zeta_1^2\\
 = 3 \Omsd ( \zeta_{\cur2} -\zeta_2 ) + 9 \Omsd
 (\zeta_{\cur1}-\zeta_1)^2
 \,,
\end{multline}
and hence, using Eqs.~(\ref{zetas2}), (\ref{defr}) and (\ref{defrSD}),
\begin{equation}
\zeta_2 = \left[ \frac{3}{2r} \left( 1+ \frac{gg''}{g^{\prime2}}
  \right) -2 -r \right] \zeta_1^2 \,.
\end{equation}
This gives the non-linearity parameter (\ref{fNLzeta}) in the sudden-decay
approximation \cite{Bartolo,LR}
\begin{equation}
\label{standardsdfNL}
\fNL = \frac{5}{4r}\left( 1+ \frac{gg''}{g^{\prime2}}
  \right) - \frac53 - \frac{5r}{6} \,.
\end{equation}

In the limit $r\to1$, when the curvaton dominates the total energy
density before it decays, we recover the non-linearity parameter
(\ref{fNLs}) of the curvaton
\begin{equation}
\fNL \to -\frac{5}{4} \left( 1 - \frac{gg''}{g^{\prime2}} \right) \,.
\end{equation}
On the other hand we may get a large non-Gaussianity ($|\fNL|\gg1$) in
the limit $r\to0$
\footnote{In this paper, by
$r \to 0$ we mean the behaviour of our expressions when $r \ll 1$.
As the \emph{fully non-linear} Eqs.~(\ref{nlzeta}) and (\ref{nlzetasimpl})
could hold even for $\zeta_\cur \gg 1$, we can, from a mathematical point
of view, consider the limit where $\zeta_\cur \to \infty$ while $\zeta$ is
kept fixed. However, as we will see later in this paper, the WMAP3 
\cite{Spergel:2006hy} upper bound for the non-linearity parameter 
$\fNL$ requires $r > 0.01$, if $g(\cur_\ast)$ is linear.
In \emph{perturbative} considerations
one should have $\zeta_{\cur 1} < 1$, and then the observed
CMB perturbation amplitude, $\zeta_1 \sim 2.5\times10^{-5}$,
would require $r \gtrsim 10^{-5}$.},
where we have
\begin{equation}
\label{fNLwhenr0}
\fNL \to \frac{5}{4r} \left( 1 + \frac{gg''}{g^{\prime2}} \right) \,.
\end{equation}

\subsection{Third order}

At third order we obtain from Eq.~(\ref{nlzetasimpl})
\begin{eqnarray}
\zeta_3 & = & \left[
\frac{9}{4r^2} \left( \frac{{{g}}^{2}{g'''}}{{g'}^3} + 3\frac{g\,{g''}}{{g'}^2}\right) 
 - \frac{9}{r} \left(1 + \frac{g\,{g''}}{{g'}^2} \right)\nonumber
   \right. \\
&& \left. + \frac{1}{2}\left( 1 - 9 \frac{g\,{g''}}{{g'}^2}\right)
+ 10 r
+ 3 r^2
\right] \zeta_1^3\,.
\label{zeta3sd}
\end{eqnarray}
The non-linearity parameter $\tauNL$ from Eq.~(\ref{tauNLzeta}) will
thus be $25/54$ times the expression in the square brackets. (As a
consistency check we note that in the limit $r \to 1$ this result
agrees with (\ref{f2s}).)

If there is non-linear evolution of the curvaton field, $\cur$,
between Hubble-exit and the start of the curvaton oscillation, such
that $gg''/{g'}^2 \simeq -1$, then from (\ref{standardsdfNL}) we see
that $\fNL$ can be small even when $r \to 0$, see also
\cite{LythNG,nurmi}. 
However, in this case $\tauNL$ will be very
large unless in (\ref{zeta3sd}) the $g^2g'''/{g'}^3$ term also
cancels the $3gg''/{g'}^2$ term. Indeed, assuming that the $g'''$
term is small, we find $\zeta_3 \to -[27/(4r^2)]\zeta_1^3$, i.e.,
$\tauNL \to -25/(8r^2)$, when $r \to 0$ if $gg''/{g'}^2 \simeq -1$.
In this situation $\zeta_3$ would be of the same order as
$\zeta_1\sim10^{-5}$, if $r \lesssim 10^{-5}$. Hence, even if the non-linear
evolution of the $\cur$ field was such that the leading order
non-Gaussianity, $\fNL$, was cancelled, the higher order terms
could still lead to large non-Gaussianity which could be ruled out by
observations.

In the absence of any non-linear evolution of $\cur$ field between
the Hubble exit and the start of curvaton oscillation (as in the
case of truly quadratic potential) we would have $g''= g''' =0$, so
the third order result would be simply
\begin{eqnarray}
\zeta_3 & = & \left[ -\frac{9}{r} + \frac{1}{2} + 10 r + 3r^2 \right]\zeta_1^3 \,,\\
\tauNL  & = & \frac{25}{54} \left[ -\frac{9}{r} + \frac{1}{2} + 10 r + 3r^2 \right] \,.
\end{eqnarray}
It should be noted that now there is no $1/r^2$ term in this
$\tauNL$.  Thus it is only at most of the same order as $\fNL$.
Indeed, in the limit $r\to 0$ we have $\fNL\to 5/(4r)$ and $\tauNL
\to -25/(6r)$, i.e., $\tauNL/\fNL \to -10/3$. As $\zeta_1 \simeq
10^{-5}$ this means that, in this case,  the third-order term
$\zeta_3$ is about 5 orders of magnitude smaller than the
second-order term $\zeta_2$ even when $r \to 0$.

\section{Numerical calculation}
\label{sectnumerical}

Although the sudden-decay approximation gives a good intuitive
derivation of both the linear curvature perturbation and the
non-linearity parameters arising from second- and third-order effects,
it is only approximate since it assumes the curvaton is not interacting with the
radiation, and hence $\zeta_\cur$ remains constant on large scales,
right up until curvaton decays. In practice the curvaton energy density is
continually decaying once the curvaton begins oscillating until
finally (when $\Gamma>H$) its density becomes negligible, and during
this decay process $\zeta_\cur$ does not remain constant
\cite{MWU,GMW}.

Another problem with results derived from the sudden-decay
approximation is that the final amplitude of the primordial
curvature perturbation, and its non-linearity, are given in terms
of the density of the curvaton at the decay time which is not
simply related to the initial curvaton density, especially as the
precise decay time, $H\sim\Gamma$, is ambiguous.

In fact a more careful treatment of the continuous decay of the
curvaton \cite{MWU,GMW} shows that the transfer coefficient at first
order, $r$ in Eq.~(\ref{defr}), is a function solely of the parameter
\begin{equation}
\label{defp}
p \equiv \left[ \Omega_\cur \sqrt{\frac{H}{\Gamma}} \right]_{\rm osc}
 \,,
\end{equation}
where the right hand side is to be evaluated when the curvaton begins
to oscillate, long before it decays, and hence can be written as
\begin{equation}
\label{peqn}
p = \frac{8\pi \sosc^2}{3M_{\rm Pl}^2} \left( \frac{m}{\Gamma}
\right)^{1/2} \,.
\end{equation}

In Refs.~\cite{MWU,GMW} the resulting primordial curvature
perturbation in the radiation-dominated era after the curvaton has
completely decayed was calculated using linear cosmological
perturbations on large scales, to give
\begin{equation}
 \label{defrofp}
 \zeta_1 = r(p) \zeta_{\cur1}
 \,,
\end{equation}
where an analytic approximation to the numerical results gives
\cite{GMW}
\begin{equation}
 \label{analyticrofp}
 r(p) \approx 1 - \left( 1+\frac{0.924}{1.24}p \right)^{-1.24} \,.
\end{equation}
We find that this not only gives a good approximation to
the amplitude of linear perturbations, but as we will show it can also be
used to give a surprisingly accurate estimate for the non-linearity
parameter $\fNL$.

In principle one could use the second-order perturbed field equations
on large scales \cite{karim} to evaluate $\zeta_2$ as a function of
$\zeta_{\cur2}$ and hence $\fNL$. Indeed this has recently been done
in Ref.~\cite{Malik:2006pm}. However a simple short-cut to the same
result is provided by the $\delta N$-formalism \cite{LR}. The
advantage of the $\delta N$-formalism is that it gives immediately the
results to any order one wants. Thus it is not necessary to repeat the
calculation with more and more complicated perturbed field equations,
if one wants results at higher order in perturbations. Indeed, once
calculated, $\delta N$ encodes all orders in perturbations, i.e., it
gives the fully non-linear $\zeta$.

\subsection{Practical implementation}

We use the evolution equations for a homogeneous Friedmann-Robertson-Walker
(FRW) universe to
describe the fully non-linear evolution in the long-wavelength limit
adopting the separate universes approach \cite{ST,WMLL}. The
resulting primordial curvature perturbation, $\zeta$, corresponds to
the perturbation in the local integrated expansion, $\delta N$, on a
final uniform-density hypersurface in the radiation-dominated
universe after the curvaton has completely decayed.
We use the fully non-linear equations for the evolution of the {\em
homogeneous} curvaton and radiation densities, including the gradual
decay of the curvaton into radiation.

Hence, our set of equations is the Friedmann equation and the
continuity equations for curvaton and radiation densities. These can
be written in the form \cite{GMW}
\begin{eqnarray}
\frac{d H_{\rm inv}}{dN} & = & \frac{3+\Omega_r}{2} H_{\rm inv} \\
\frac{d \Omega_\cur}{dN} & = &  \Omega_\cur \Omega_r - \Gamma \Omega_\cur H_{\rm
  inv} \\
\frac{d \Omega_r}{dN} & = &  \Gamma \Omega_\cur H_{\rm inv} + \Omega_r(\Omega_r-1)\,,
\end{eqnarray}
which is particularly suitable for numerical calculation. Here
$H_{\rm inv} \definit 1/H$, $\Omega_\cur = \rho_\cur/\rho_{\rm tot}$, and
$\Omega_r = \rho_r / \rho_{\rm tot}$ with $\rho_{\rm tot} = \rho_\cur + \rho_r$.

Since the end-result does not depend on a particular choice of
$\Gamma$ and $m$, as long as we integrate far enough so that the
curvaton has completely decayed at the end of calculation, we fix
these to the values $m=10^{-5} M_{\rm Pl}$ and $\Gamma = 10^{-7} m$.
The initial value of $H_{\rm inv}$ is $1/m$, since we start the
calculation at the beginning of the curvaton oscillation. After
specifying the value of $p$, we calculate the initial values of
$\Omega_\cur$ and $\Omega_r$. They are $\Omega_{\cur i} = (\Gamma
H_{\rm inv})^{1/2}\,p$, and $\Omega_{ri} = 1 - \Omega_{\cur i}$. The
initial value for our integration variable $N$ can be set to zero
because in the absence of any initial perturbation in the radiation
($\zeta_r=0$) the initial surface is both spatially flat and
has uniform energy density and Hubble rate $H=H_i=m$ (recall that
from the Friedmann equation $\rho_{\rm tot}\propto H^2$). We are
interested in the integrated expansion between this initial
unperturbed hypersurface and some final uniform-density surface
$H=H_f\ll\Gamma$.  Then the (local) integrated expansion between
these surfaces will be just the final value of $N=N_f$.

We find that $\Omega_\cur$ is practically zero when $N \gtrsim 11$.
From this we deduce that a suitable ending condition (curvaton has
completely decayed) is $H=H_f = \Gamma/500$. We use our modified
version of an adaptive step size ode integrator
\cite{NumericalRecipes} and the accuracy
parameter ${\rm eps} =10^{-21}$. We start integration with a
sufficiently small step size as demanded for our required accuracy.
Finally, when $N$ starts to be of the order $11$ the step trial
would lead to $H_{\rm inv} > 1/H_f$. As soon as this happens we
divide the step trial by two. We repeat this procedure until $H_{\rm
inv}$ obeys $(1-10^{-20})/H_f < H_{\rm inv} < 1/H_f$. Then we save
$p$ and the final $N_f$. To find $N(p)$ we repeat this process for
50000 logarithmically spaced values of $p$ in the range $[10^{-8},\,
10^4]$.


\subsection{Comparison of sudden-decay approximation with numerical results}

Previous studies of non-Gaussianity in the curvaton model have been
based on the sudden-decay approximation. In \cite{Moriond} we
extended the calculation of the non-linearity parameter $\fNL$ to
the non-instantaneous decay case and found that the sudden-decay
approximation is indeed very accurate. Recently, a similar numerical
comparison was made in \cite{Malik:2006pm} using second order
perturbation theory. Our results obtained using $\delta N$ formalism
agree with those of Ref.~\cite{Malik:2006pm}. In this subsection we
describe our calculation of $\fNL$ \cite{Moriond} in more detail and
for the first time perform similar studies for $\tauNL$.

Expanding $\delta N$ we have
\begin{equation}
 \label{zetaexpansion}
\zeta = N'\delta\cur_\ast
+ \half N''(\delta\cur_\ast)^2
+ \textstyle\frac{1}{6} N''' (\delta\cur_\ast)^3
+ \ldots
\end{equation}
Comparing this with Eq.~(\ref{zetanl}) we can read off
$\zeta_n=\partial^n N/\partial \chi_*^n$. Substituting this into
(\ref{fNLzeta}) gives \cite{LR}
\begin{eqnarray}
\label{fNLdeltaN}
\fNL   & = & \frac{5}{6}\frac{N''}{{N'}^2} \,,
\end{eqnarray}
and substituting into (\ref{tauNLzeta}) we find
\begin{eqnarray}
\label{tauNLdeltaN}
\tauNL & = & \frac{25}{54} \frac{N'''}{{N'}^3} \,.
\end{eqnarray}

As
we will specify the initial conditions for our numerical solutions
by giving the value of $p$, defined in Eq.~(\ref{defp}), the
differentiations of $N$ with respect to $\cur_\ast$ need to be
converted into differentiations with respect to $p$. From
(\ref{peqn}) we have
\begin{equation}
\label{asttop}
\frac{\partial}{\partial\cur_\ast} = 2p\frac{g'}{g}\frac{\partial}{\partial p} \,.
\end{equation}
Using this we find
\begin{eqnarray}
\label{Nprimeofp}
  N' & = &\!\!
2 p \frac{g'}{g} \dNi \,,\\
\label{Nprime2ofp}
 N'' & = &\!\!
4p^2 \left(\frac{g'}{g}\right)^2 \dNii
+ 2p \left[ \left(\frac{g'}{g}\right)^2 \!  + \! \frac{g''}{g} \right]\dNi
\,,\\
\label{Nprime3ofp}
  N''' & = &\!\!
8p^3 \left(\frac{g'}{g}\right)^3 \dNiii
+ 12p^2 \left[ \frac{g'g''}{g^2} + \left(\frac{g'}{g}\right)^3 \right]\dNii \nonumber\\
&& + 2p \left[ \frac{g'''}{g} + 3 \frac{g'g''}{g^2} \right]\dNi \,.
\end{eqnarray}
Recalling the definition (\ref{defrofp}) of the curvature
perturbation transfer efficiency at linear order, $r$, we find, from
Eqs.~(\ref{zetas1}) and ({\ref{Nprimeofp}),
\begin{equation}
\label{rofp}
r = 3p\dNi \,.
\end{equation}

Once we have numerically calculated $N$ as a function of $p$ in the
non-instantaneous decay case, we can calculate $\fNL(r)$ and
$\tauNL(r)$ by substituting (\ref{Nprimeofp}) -- (\ref{rofp}) into
(\ref{fNLdeltaN}) and (\ref{tauNLdeltaN}). For example, for $\fNL$
we find
\begin{equation}
\label{fNLdndpform}
\fNL = \frac{5}{4r}\left( 1+ \frac{gg''}{g^{\prime2}}\right)
+ \frac{5}{6} \frac{d^2N/dp^2}{\left( dN/dp\right)^2} \,.
\end{equation}
Comparing to the sudden-decay result (\ref{standardsdfNL}) we see
that the first term is exactly the same. Numerical calculation of
the derivatives of $N$ with respect to $p$ shows that the second
term approaches a constant value $-2.27$ as $r\to 0$. In the
sudden-decay case it approaches a constant $-5/3=-1.67$, see
(\ref{standardsdfNL}). Thus in both cases
\begin{equation}
\fNL \to \frac{5}{4r}\left( 1+ \frac{gg''}{g^{\prime2}}\right)
\end{equation}
when $r \to 0$. In the opposite limit, $r\to 1$, both results give
$\fNL \to -5/4$. So any difference between the sudden-decay
approximation and the non-instantaneous decay calculation appears
only at intermediate values of $r$, i.e., when the radiation energy
density from the inflaton decay products and the curvaton energy
density are of the same order when the curvaton decays, $\rho_{r,
\rm dec} \sim \rho_{\cur, \rm dec}$.

The second term in (\ref{fNLdndpform}) can also be written in terms
of $r$. Using (\ref{rofp}) and (\ref{asttop}) the result
(\ref{fNLdndpform}) reads
\begin{equation}
 \label{fNLrprime}
\fNL = \frac{5}{4r}\left( 1+ \frac{gg''}{g^{\prime2}}\right) +
\frac{5}{4} \frac{(g/g')r' - 2r}{r^2} \,.
\end{equation}
where we have used
\begin{equation}
 \label{rprime}
r' = 2 \frac{g'}{g} \left( r + 3p^2 \frac{d^2 N}{dp^2} \right) \,.
\end{equation}

Comparing (\ref{fNLrprime}) to (\ref{standardsdfNL})
we find that in the sudden-decay case
\begin{equation}
\label{rprimeSD} \frac{g}{g'}r_{\rm SD}' = 2r_{\rm SD} -
\frac{4}{3}r_{\rm SD}^2 - \frac{2}{3}r_{\rm SD}^3 \,,
\end{equation}
whereas in the non-instantaneous case $r'$ must be determined
numerically. Thus one way to characterise the accuracy of the
sudden-decay approximation is to calculate $r'$ numerically,
employing (\ref{rofp}) and (\ref{rprime}),
and compare it to the above expression (\ref{rprimeSD}) in the
sudden-decay approximation.

\begin{figure}
\vspace{1mm}
\begin{center}
\includegraphics[width=0.45\textwidth]{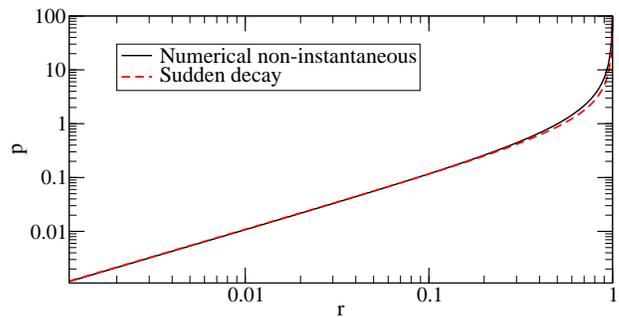}
\end{center}
\caption{To achieve the same curvature perturbation
transfer efficiency $r$ one needs to start from
slightly different initial value of $p$
in the sudden-decay case ({\it red dashed line})
than in the non-instantaneous decay
case ({\it black solid line}).
\label{Fig:pvsr}}
\end{figure}

As mentioned in the beginning of this section the relation between
$r$ and $p$ in the sudden-decay approximation is non-trivial. In the
non-instantaneous decay case $r(p)$ is easy to find numerically from
(\ref{rofp}). In the sudden-decay case we can only determine $r_{\rm
SD}(p)$ from (\ref{defrSD}) if we know $\Omega_{\cur,{\rm
dec}}(p)$.

Fortunately, a short-cut to the same result is provided by the
differential equation (\ref{rprimeSD}). Using (\ref{asttop}) we find
from (\ref{rprimeSD})
\begin{equation}
\int \frac{dp}{p} = \int \frac{3dr_{\rm SD}}{3r_{\rm SD} - 2r_{\rm
SD}^2 - r_{\rm SD}^3} \,,
\end{equation}
and hence
\begin{equation}
p \propto r_{\rm SD} (r_{\rm SD}+3)^{-1/4} (1-r_{\rm SD})^{-3/4} \,.
\end{equation}
The constant of proportionality is not uniquely determined by the
sudden-decay approximation, and corresponds to the arbitrariness in
the definition of the decay time $H_{\rm dec}\sim \Gamma$.

What we can do is to use the limiting form of the analytic
approximation to the numerical solution (\ref{analyticrofp}) for
small $p$ to provide an overall normalisation for the sudden-decay
approximation. This yields
\begin{equation}
\label{ourprsd}
p = \frac{3^{1/4}r_{\rm SD}}{0.924 (r_{\rm SD}+3)^{1/4} (1-r_{\rm
SD})^{3/4}} \,.
\end{equation}
The above equation thus determines the value of $p$ that
corresponds to a given value of the linear transfer function,
$r_{\rm SD}$, in the sudden-decay approximation, and hence
$\Omega_{\cur,{\rm dec}}$ from Eq.~(\ref{defrSD}).
In Fig.~\ref{Fig:pvsr} we show $p$ as a function of $r_{\rm SD}$
in the sudden-decay approximation and compare this with the
numerical non-instantaneous decay result for $p(r)$.

Our form for $r_{\rm SD}(p)$ is quite different from that adopted by
Malik and Lyth in their recent work \cite{Malik:2006pm}. They used a
much simpler, but less accurate, estimate for $r(p)$ in the
sudden-decay approximation:
\begin{equation}
 r_{\rm ML} = \frac{p}{1+p} \,.
\end{equation}
Although $r_{\rm ML}\to1$ as $p\to\infty$ it fails to reproduce the
correct linear coefficient as for $p\to0$.
The apparent error in the sudden-decay approximation for $\fNL$
reported by Malik and Lyth \cite{Malik:2006pm} (see for instance
Figure~9 in that paper) is primarily due to this inaccuracy in the
$r_{\rm ML}(p)$.


In what follows we have chosen to do all our comparisons of the
sudden-decay approximation and numerical non-instantaneous decay
results at common values of $r$. In other words, we have presented
our results as a function of $r$ instead of $p$.
%
%
After taking into account this fundamental difference in thinking,
the results of \cite{Malik:2006pm} agree with ours.
%

Now we are ready for the final comparison of ${\fNL}_{\rm SD}(r)$,
derived in the sudden-decay approximation with, ${\fNL}(r)$ calculated
numerically allowing for non-instantaneous decay. Fig.~\ref{Fig:fNL}
shows that if $\fNL>60$ ($r<0.02$) or $\fNL<-1.16$ ($r>0.95$),
the sudden decay result differs from the non-instantaneous decay
result by less than $1\%$. Hence, when constraining the curvaton model
with the current observational constraints on $\fNL$ there is no need
for an exact numerical calculation; using the sudden-decay
approximation is sufficient. However, in the future experiments are
expected to bring down the upper bound on $|\fNL|$, and then
constraining the curvaton model does require the numerical
calculation presented here (or in \cite{Moriond,Malik:2006pm}).

\begin{figure}
\begin{center}
\includegraphics[width=0.45\textwidth]{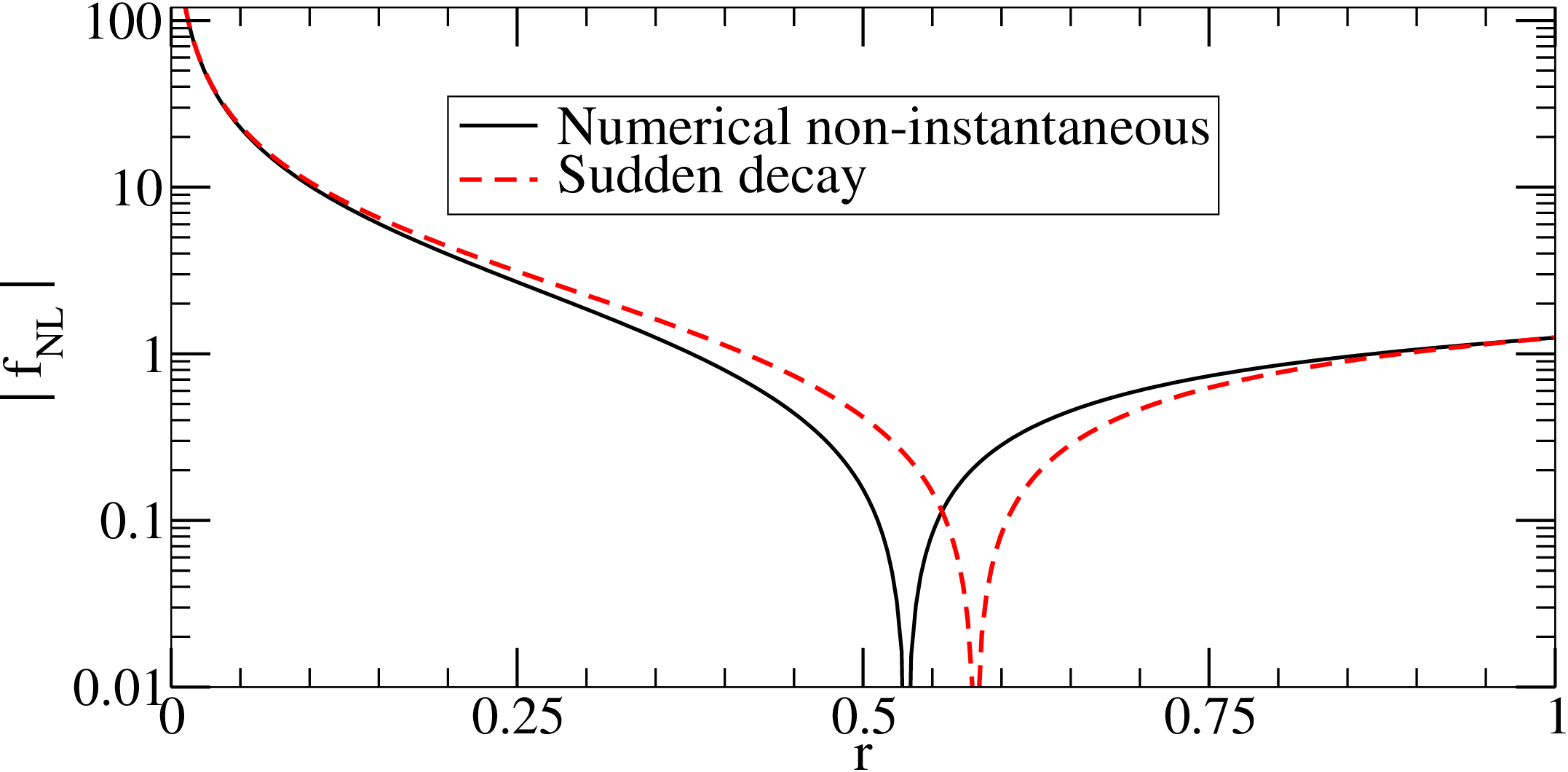}
\end{center}
\caption{The non-linearity parameter $\fNL$ as a function of curvature
perturbation transfer efficiency $r = \zeta_1 / \zeta_{\cur 1}$.
The analytical approximative, i.e., sudden-decay result ({\it red dashed line})
crosses zero at $r = 0.58$ and is negative for $r>0.58$. The exact numerical
result ({\it black solid line}) is negative for $r>0.53$. 
Here we assume that $g(\cur_\ast)$ is linear.
\label{Fig:fNL}\vspace{2mm}}
\end{figure}

\begin{figure}
\begin{center}
\includegraphics[width=0.47\textwidth]{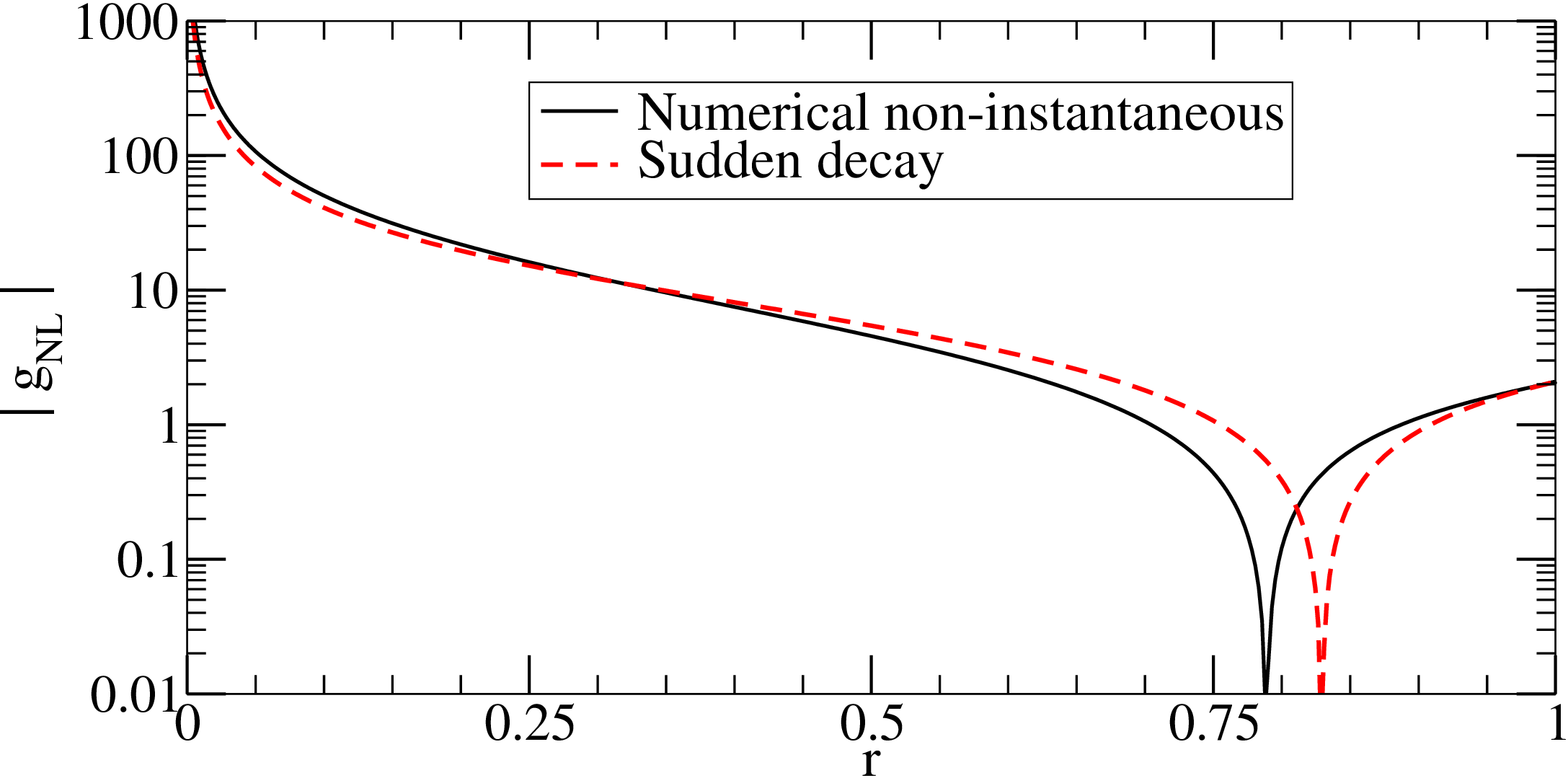}
\end{center}
\caption{The non-linearity parameter $\tauNL$ as a function of curvature
perturbation transfer efficiency $r = \zeta_1 / \zeta_{\cur 1}$.
The analytical approximative, i.e., sudden-decay result ({\it red dashed line})
crosses zero at $r = 0.83$ and is positive for $r>0.83$. The exact numerical
result ({\it black solid line}) is positive for $r>0.79$.
Here we assume that $g(\cur_\ast)$ is linear.
\label{Fig:tauNL}}
\end{figure}

In Fig.~\ref{Fig:tauNL} we compare $\tauNL$ in the sudden-decay
approximation with the non-instantaneous decay result. The sudden-decay
result for $\tauNL$ is much more inaccurate than for $\fNL$.
However, the present observational constraints on $\tauNL$
are so weak that again the sudden-decay approximation may be sufficient.

Let us end this subsection with a comment on numerical accuracy.
Since the derivatives of $N(p)$ in (\ref{Nprimeofp}) --
(\ref{Nprime3ofp}) involve subtraction of nearly equal numbers, the
calculation must be carried out carefully. The first requirement is
that the integration step-size in $N$ is small enough compared to
$\delta p$ that the two initial values $p_i$ and $p_{i+1}= p_i
+\delta p$ really lead to numerically different values for
$N_i=N(p_i)$ and $N_{i+1}=N(p_{i+1})$. The accuracy of $N$ must be
good enough to maintain enough significant figures in $\delta N =
N_{i+1} - N_i$. The smaller the steps in $\delta p$ that we want to
take the higher the accuracy in $N$ that we need. We calculate the
first derivative at $p=p_i$ as an average of two nearby gradients
\begin{equation}
\left.\dNi\right|_{p=p_i}  = \frac{1}{2}\left(
  \frac{N_i-N_{i-1}}{p_i-p_{i-i}} + \frac{N_{i+1} - N_i}{p_{i+1}-p_i} \right) \,.
\end{equation}
We use the same algorithm for the second derivative (with $N$
replaced by the result of the calculation of $dN/dp$) and for the
third derivative (with $N$ replaced by the result of the calculation
of $d^2N/dp^2$). As a result, the first derivative picks up
contributions from 3 nearby points, the second derivative picks up
weighted contributions from 5 nearby points and the third derivative
from 7 nearby points. This procedure smooths out any residual
numerical noise.


\subsection{An analytic approximation to the numerical result}

The analytic approximation of $r(p)$, Eq.~(\ref{analyticrofp}), with
help of (\ref{asttop}) gives
\begin{equation}
\label{eq:rfitder}
  r' \approx
2\times 1.24 \left( 1-r \right)\left[ 1 - (1 - r)^{\frac{1}{1.24}}
\right] \frac{g'}{g}\,.
\end{equation}
Hence, from (\ref{fNLrprime}) we find an analytic approximation to
the non-linearity parameter
\begin{eqnarray}
\label{eq:fNLfit}
  &&{\fNL}_{\rm fit} =
 \frac{5}{4} \frac{1}{r} \left( 1 + \frac{g g''}{{g'}^2} \right)
\\
 & &\quad
 +\frac{5}{4} \frac{1}{r^2}\left\{-2r + 2\times
 1.24 \left( 1-r \right) \left[ 1 - (1 - r)^{\frac{1}{1.24}} \right]\right\}\,.
 \nonumber
\end{eqnarray}
The difference from the numerical result  is non-negligible only when
$\fNL$ is extremely close to zero. Indeed, we find
\begin{equation}
  \left|\frac{{\fNL}_{\rm fit} - \fNL}{\fNL}\right| < 1\%\,
\end{equation}
if $r<0.501$ or $r>0.542$. The difference is larger than 5\% only
when $r\in[0.528,\, 0.534]$.

\section{Probability density function}
\label{sectpdf}

Thus far we have calculated the second- and third-order corrections
to the curvature perturbation produced by the curvaton decay from
which the leading order terms to the bispectrum and trispectrum can
be calculated. However the $\delta N$-formalism allows us to
describe the full non-linear probability density function on
large scales for the non-linear primordial curvature perturbation
defined in Eq.~(\ref{zetanl}).

Assume we have two random variables $y$ and $z$, and the pdf of $y$
is $f(y)$. Furthermore, assume that the functional dependence of $z$
on $y$ is known, $z=z(y)$, and this mapping is a bijection. Then the
probability of $z$ being in the interval $(z_1, z_2)$ is given by
\begin{equation}
  P(z_1 < z < z_2)
  = \int_{z_1}^{z_2} \left|\frac{dy}{dz}\right| f(y) dz\,,
\end{equation}
where the absolute value is needed in the case that $y(z)$ happens
to be a decreasing function. Hence the pdf of the random variable $z$
is
\begin{equation}
  \tilde f(z) =  \left|\frac{dy(z)}{dz}\right| f[y(z)] \,.
\end{equation}
In the multi-variable case the derivative would be replaced by the
Jacobian determinant.
For a Gaussian random variable, $y$, with mean $\mu_y$ and variance
$\sigma_y^2$ we have
\begin{equation}
  f(y) = \frac{1}{\sqrt{2\pi\sigma_y^2}}e^{-(y-\mu_y)^2/(2\sigma_y^2)} \,.
\end{equation}

%

Since the first-order primordial curvature perturbation, $\zeta_1$,
depends linearly on the initial Gaussian field perturbation,
$\delta\cur_\ast$, we can take $\zeta_1$ as our Gaussian
``reference'' variable with mean $\mu_{\zeta_1}=0$ and variance
\begin{equation}
 \sigma_{\zeta_1} = \frac{2}{3} r \frac{g'}{g} \sigma_{\delta\cur_*} \,.
\end{equation}
With this goal in our mind we have already written all our
non-linear expressions for $\zeta$ as a function of $\zeta_1$. In
the sudden-decay approximation we found an analytic functional
dependence $\zeta=\zeta(\zeta_1)$ and in the non-instantaneous decay
case this function was found numerically. The mapping from $\zeta_1$
to $\zeta$ is not actually a bijection as there can be several
values of $\zeta_1$ which are mapped onto the same value of $\zeta$.
See Appendix A for the sudden-decay case. Calling these values
${\zeta_1}_j$, it is now easy to calculate the pdf of the non-linear
primordial curvature perturbation
\begin{equation}
\label{fulltildepdf}
  \tilde f(\zeta) = \sum_j \left| \frac{d\zeta_1}{d\zeta}
  \right|_{\zeta_1={\zeta_1}_j} \! \! \! f_g({\zeta_1}_j) \,,
\end{equation}
where $f_g(\zeta_1)$ it the Gaussian pdf with $\mu = 0$ and $\sigma
= \sigma_{\zeta_1}$. 

For simplicity we will assume in the rest of this section that there
is no non-linear evolution of the curvaton field before it begins to
oscillate, so that $g \propto \cur_\ast$, i.e., $g^{(n)}=0$ for
$n\geq2$. In principle one could also carry through the non-linear
evolution of the curvaton field into the full numerical calculation of
the pdf for the primordial curvature perturbation.

At the end of Appendix A we derive $\tilde f(\zeta)$ in the
sudden-decay approximation, see Eq.~(\ref{fullsdpdf2}).
Here we continue by demonstrating the calculation of $\tilde
f(\zeta)$ up to second order, which we will call $\tilde
f_2(\zeta)$. {}From (\ref{fNLtauNL}) we have $\zeta = \zeta_1 +
\frac{3}{5}\fNL\zeta_1^2$ up to second order, i.e.,
\begin{equation}
  {\zeta_1}_{\pm} = \frac{5}{6\fNL} \left(- 1  \pm \sqrt{1+12\fNL\zeta/5} \right)\,.
\end{equation}
Substituting this into (\ref{fulltildepdf}) we find
\begin{multline}
\label{pdf2order}
  \tilde f_2(\zeta)  =
\frac{1}{\sqrt{1+12\fNL\zeta/5}}
\frac{1}{\sqrt{2\pi\sigma_{\zeta_1}^2}} \\
\times \sum_{\pm}
e^{-\left[\frac{5}{6\fNL} \left(- 1  \pm \sqrt{1+12\fNL\zeta/5}
    \right)\right]^2
\Big/ (2\sigma_{\zeta_1}^2)}\,,
\end{multline}
if $\zeta > -5/(12\fNL)$, and $\tilde f_2(\zeta)=0$ otherwise. If we
want to evaluate $\tilde f_2(\zeta)$ for a particular value of the
parameter $r$ we can substitute $\fNL$ from (\ref{standardsdfNL})
into (\ref{pdf2order}) in the sudden-decay case, or use $\fNL(r)$ as
shown in Fig.~\ref{Fig:fNL} in the non-instantaneous decay case.

In Fig.~\ref{Fig:pdfcomparisonLARGEfnl}(a) we compare the fully
non-linear pdf $\tilde f(\zeta)$ (non-instantaneous decay) and $\tilde
f_{\rm SD}(\zeta)$ (sudden decay) to the Gaussian $f_g(\zeta_1)$ in the case
when there is large non-Gaussianity ($r=0.00028$, $p=0.00030$,
$\fNL=4432$). In Fig.~\ref{Fig:pdfcomparisonWMAP}(a) we plot the pdfs
in the case when $\fNL$ has exactly the WMAP3 upper limit value
($r=0.010758$, $p=0.011560$, $\fNL=114$). When the
non-linearity parameter is very large, this kind of visual comparison
reveals the non-Gaussianity, but already with $\fNL=114$ the $\tilde
f$ is virtually indistinguishable from the Gaussian $f_g$. However, in
Fig.~\ref{Fig:pdfcomparisonLARGEfnl}(b) and
Fig.~\ref{Fig:pdfcomparisonWMAP}(b) we plot $\tilde f/f_g$ which
reveals the non-Gaussianity even when $\fNL=114$.

\subsection{Moments of the distribution}

The non-Gaussianity can be described quantitatively by calculating
the moments of the pdf. Any pdf $f(z)$ should give a unit total
probability
\begin{equation}
  \int f(z) dz = 1\,.
\end{equation}
The mean can be calculated as
\begin{equation}
  \mu_z = \int z f(z) dz \,.
\end{equation}
and the $i^{\rm th}$ moment $m_z(i)$ is defined as
\begin{equation}
\label{momentdef}
  m_z(i) = \int (z-\mu)^i f(z) dz\,.
\end{equation}
The second moment is the variance ($\sigma_z^2$), the third moment is
called skewness, and the fourth moment kurtosis. As these moments can
be extracted from the CMB maps it is enlightening to calculate the
curvaton-model prediction for them.

\begin{figure}
\begin{center}
\includegraphics[width=0.45\textwidth]{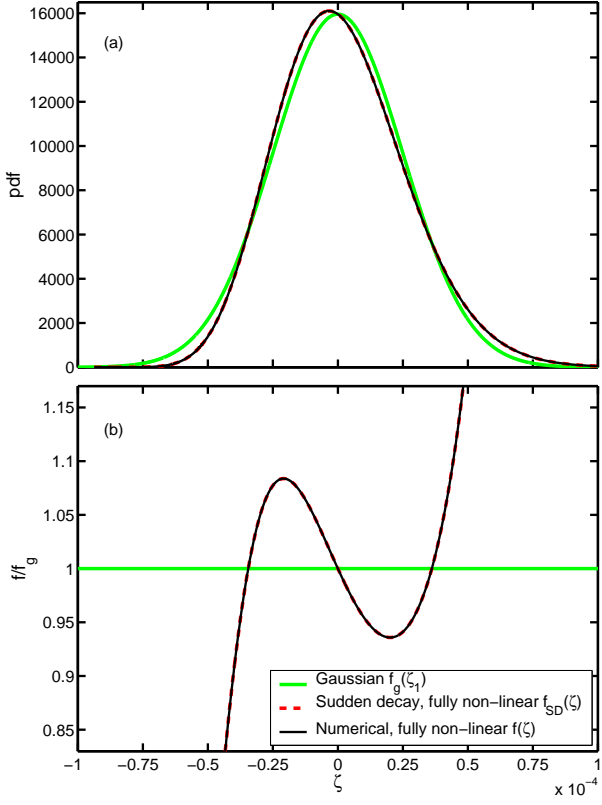}
\end{center}
\caption{(a) Pdfs at $r=0.00028$ ($p=0.00030$, $\fNL=4432$).
Red dashed line is for the sudden decay, $\tilde f_{\rm SD}(\zeta)$,
and black solid line for the non-instantaneous decay, $\tilde f(\zeta)$.
The solid green/grey line is the Gaussian ``reference'', $f_g(\zeta_1)$.
(b) The ratio of non-Gaussian pdfs to the Gaussian one.
\label{Fig:pdfcomparisonLARGEfnl}}
\end{figure}
\begin{figure}
\begin{center}
\includegraphics[width=0.45\textwidth]{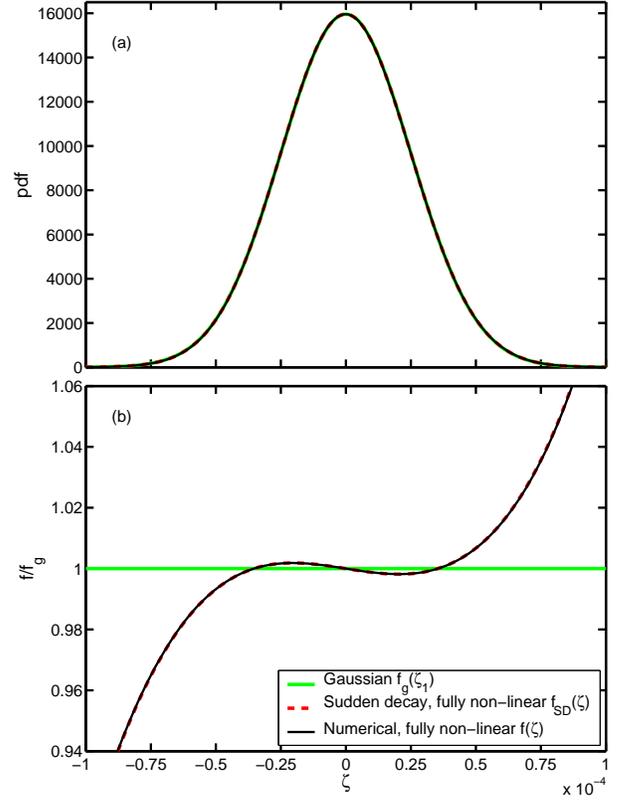}
\end{center}
\caption{(a) Pdfs as in Fig.~\ref{Fig:pdfcomparisonLARGEfnl} but now 
 at $r=0.010758$ ($p=0.011560$, $\fNL=114$). 
 In this figure the Gaussian reference (solid green/grey line)
 is completely indistinguishable from the fully non-linear
 (``non-Gaussian'') pdfs.
(b) The ratio of non-Gaussian pdfs to the Gaussian one.
\label{Fig:pdfcomparisonWMAP}}
\end{figure}

For Gaussian pdfs any odd moment (with $i \ge 3$) is zero, since the
probability density is symmetric around the mean. The even moments
of a Gaussian distribution are easy to calculate employing partial
integration to give $m(4) = 3\sigma^4$, $m(6) = 15\sigma^6$, $m(8) =
105\sigma^8$, $m(10) = 945\sigma^{10}$, $m(12) = 10395\sigma^{12}$, 
etc. Any departure from these values indicates that
the pdf is non-Gaussian. If odd moments differ from zero, there is
an asymmetric deviation from Gaussianity. If even moments are
smaller than in the Gaussian case, the pdf is more sharply peaked
than the Gaussian. If even moments are larger, the pdf is wider. The
set of moments $\{ \mu, m(i) | i=2...\infty\}$ encodes the same
information of non-Gaussianity as our fully non-linear
$\zeta(\zeta_1)$ (or the expansion $\zeta = \sum_{n=1}^\infty
\zeta_n/n!$). However, it should be noted that giving the value,
for example, for $m_\zeta(3)$ is not simply equivalent to giving the
value for $\fNL$, because the moment picks up contributions from the
fully non-linear $\zeta$, not just from
$\zeta_1+\frac{3}{5}\fNL\zeta_1^2$.

It turns out that the moments can be calculated very accurately
using the $\delta N$-formalism, even in the non-instantaneous decay
case, since we need not to calculate the derivatives of the local
expansion, $N$, as was done in calculating $\fNL$ or $\tauNL$. At first
it seems that we would need $\tilde f(\zeta)$, which includes a
numerical derivative $d\zeta_1/d\zeta$:
\begin{equation}
  m_\zeta(i) = \int (\zeta-\mu)^i \tilde f(\zeta) d\zeta\,,
\end{equation}
but substituting $\tilde f$ from Eq.~(\ref{fulltildepdf})
we end up with
\begin{equation}
\label{non-linearmoments}
  m_\zeta(i)  =
 \sum_j \int (\zeta-\mu)^i f_g(\zeta_{1j}) d\zeta_{1j} \,.
\end{equation}
Here the only numerically calculated quantity is $\zeta$
(and $\mu$).

\begin{figure*}
\begin{center}
\includegraphics[width=0.8\textwidth]{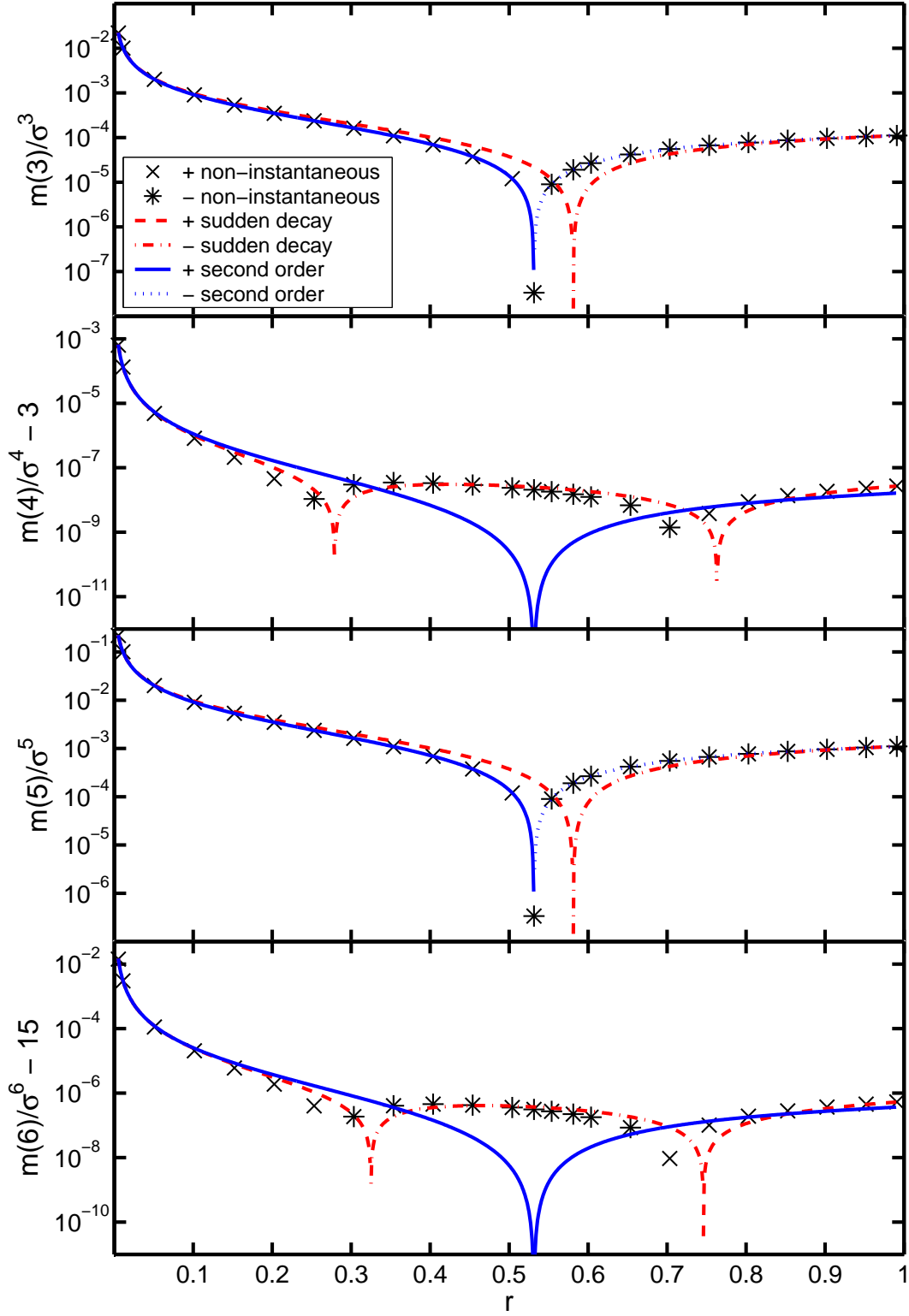}
\end{center}
\caption{Third, fourth, fifth and sixth (from top to bottom) moments of the probability
  density function of the primordial curvature perturbation $\zeta$ 
  as a function of the linear transfer parameter, $r$. 
  The predictions using the fully non-linear
  sudden-decay approximation are shown by the dashed red
  line (dot-dashed for negative values),
  and the second-order results are shown by the solid blue line 
  (dotted for negative values).
  The fully non-linear numerical results
  are indicated by black crosses (stars for negative values).
\label{Fig:moments}}
\end{figure*}

Calculating the moments in the second order expansion and comparing
to the results of fully non-linear calculation
(\ref{non-linearmoments}) we can address the question whether
$\sum_{n=3}^\infty \zeta_n/n!$ gives an important contribution to
$\zeta$ and hence to the non-Gaussianity or whether the second order
expansion $\zeta \approx \sum_{n=1}^2 \zeta_n/n!$ is indeed
accurate enough. To this end let us calculate in the second-order
expansion the mean
\begin{eqnarray}
 \label{m2mean}
\mu_2 & = &\int \zeta f_g(\zeta_1) d\zeta_1
= \int \left( \zeta_1 +\frac{3}{5}\fNL\zeta_1^2 \right) f_g(\zeta_1) d\zeta_1 \nonumber\\
& = & \frac{3}{5}\fNL\sigma^2_{\zeta_1} \,,
\end{eqnarray}
the variance
\begin{eqnarray}
\sigma_2^2 & = &
\int \left( \zeta_1 +\frac{3}{5}\fNL\zeta_1^2  -
\frac{3}{5}\fNL\sigma^2_{\zeta_1}\right)^2 f_g(\zeta_1) d\zeta_1 \nonumber\\
& = & \sigma^2_{\zeta_1} + 2 \left( \frac{3}{5}\fNL \right)^2\!\!
\sigma^4_{\zeta_1} \,,
\end{eqnarray}
the skewness
\begin{eqnarray}
\label{m2skewness}
m_2(3)
& = & 6 \left(\frac{3}{5}\fNL \right) \sigma^4_{\zeta_1} + 8 \left( \frac{3}{5}\fNL \right)^3\!\!
\sigma^6_{\zeta_1} \,,
\end{eqnarray}
and the kurtosis
\begin{eqnarray}
 \label{m2kurtosis}
 \! \! \! \! \! \! m_2(4) \!  & = &
\!  3\sigma^4_{\zeta_1}
\! + \! 60  \left( \frac{3}{5}\fNL \right)^2\!\! \sigma^6_{\zeta_1}
\! + \! 60  \left( \frac{3}{5}\fNL \right)^4\!\! \sigma^8_{\zeta_1}\!.
\end{eqnarray}
For higher moments in the second-order expansion we find
\begin{eqnarray}
 \label{m25}
 m_2(5)  & = &
60 \left(\frac{3}{5}\fNL \right) \sigma^6_{\zeta_1} + 680 \left( \frac{3}{5}\fNL \right)^3 \sigma^8_{\zeta_1}\nonumber\\
 &+& 544 \left( \frac{3}{5}\fNL \right)^5 \sigma^{10}_{\zeta_1}\,,
%
\end{eqnarray}
and
\begin{eqnarray}
 \label{m26}
 m_2(6) & = &
15\sigma^6_{\zeta_1}
+ 1170  \left( \frac{3}{5}\fNL \right)^2 \sigma^8_{\zeta_1}\nonumber\\
&+& \!\! 9060 \! \left( \frac{3}{5}\fNL \right)^4 \!\!\!\sigma^{10}_{\zeta_1}
+ 6040 \! \left( \frac{3}{5}\fNL \right)^6 \!\!\!\sigma^{12}_{\zeta_1}\,.
%
%
%
\end{eqnarray}
From Eqs.~(\ref{m2skewness}) and (\ref{m25}) we find the leading order
prediction
\begin{equation}
\label{lopred53}
\frac{m_2(5)/\sigma^5}{m_2(3)/\sigma^3} = 10\,,
\end{equation}
and from Eqs.~(\ref{m2kurtosis}) and (\ref{m26}) we expect
\begin{equation}
\label{lopred64}
\frac{m_2(6)/\sigma^6-15}{m_2(4)/\sigma^4-3} \approx \frac{1170}{60} = 19.5\,,
\end{equation}
at least when $\fNL$ is small.

In Fig.~\ref{Fig:moments} we plot the moments from the third up to the
sixth one as a function of $r$. (For each value of $r$, it takes
couple of hours to calculate the moments in the non-instantaneous
decay case with our code on a typical PC. This comes about because we
want to find $N$ with a relative error of less than $10^{-20}$ in
order to have a sufficiently accurate result near to the peak of the
pdf where $\zeta=\delta N$ is extremely close to zero.
We need about $2\times10^5$ steps integrating $N$ from the Friedmann
equation, and we calculate $\zeta$ for $6001$ equally spaced values
of $\zeta_1$ in the range $[-6\times10^{-3},\,6\times10^{-3}]$.
Thus to produce a pdf for a fixed value of $r$ we need $10^9$
integration steps.) We compare the fully non-linear
non-instantaneous decay result to the second-order results and to the
fully non-linear sudden-decay result. 

%


We find that results obtained using the fully non-linear
sudden-decay approximation
agree well with those obtained from the full non-instantaneous decay.
The sudden-decay approximation accurately predicts the moments of the
distribution for small values of $r$ where the non-Gaussianity is
largest, and only fails to give the precise values of $r$ where the
moments cross zero, very similar to what was seen previously when evaluating
the non-linear parameters $\fNL$ and $\tauNL$ in Sec.~IV. 

The expressions for the moments, given in
Eqs.~(\ref{m2mean}--\ref{m26}), calculated using only terms up
to second-order in perturbation theory [but using the full numerical
value for $\fNL(r)$] are an excellent description of the odd moments
of the distribution for all $r$. For even moments, the
second-order expressions give the correct order magnitude, but cannot
always reproduce the correct sign of the moments for $r>0.1$. In
particular we see that the even moments of the distribution predicted
at second-order are always larger than the Gaussian value (setting
$\fNL=0$), whereas the full numerical results show that the even
moments can be less than the Gaussian value. To describe the
deviations from Gaussianity in the even moments we need to include
third-order terms. For example the variance, to third order, is given
by
\begin{eqnarray}
\sigma_3^2 
& = & \textstyle \sigma^2_{\zeta_1} + \left[2\left( \frac{3}{5}\fNL \right)^2 + 6 \left( \frac{9}{25}\tauNL \right)\right]
\sigma^4_{\zeta_1}\nonumber\\
& + & \textstyle 15 \left( \frac{9}{25}\tauNL \right)^2 \sigma^6_{\zeta_1}\,,
%
\end{eqnarray}
where the expression in square brackets gives the leading order correction to the
Gaussian result. This correction
can be negative due to negative $\tauNL$ when $\fNL$ is small. On the contrary,
the skewness up to third order is
\begin{multline}
m_3(3) = \textstyle 6 \left(\frac{3}{5}\fNL \right) \sigma^4_{\zeta_1} + \Big[8 \left( \frac{3}{5}\fNL \right)^3
 + 72  \left(\frac{3}{5}\fNL \right)\\
\times \textstyle \left(\frac{9}{25}\tauNL \right)\Big] \sigma^6_{\zeta_1}
 + 270 \left(\frac{3}{5}\fNL \right) \left(\frac{9}{25}\tauNL \right)^2 \sigma^8_{\zeta_1}\,,
%
\end{multline}
where the first term is the leading order correction to the
Gaussian result [$m(3)=0$]. As seen this correction does not have
$\tauNL$ term so that already the second order expansion leads to
approximately correct results.

Although not shown in the Fig.~\ref{Fig:moments} we have verified
that the third-order perturbation theory (using the numerical results
for $\fNL$ and $\tauNL$) accurately reproduce all the moments of the
distribution as a function of $r$ at least up to and including the
sixth moment.

The results (see Fig.~\ref{Fig:moments}) obey the predictions
of Eqs.~(\ref{lopred53}) and (\ref{lopred64}). In particular,
the fully non-linear numerical calculation
also reproduces the predicted ratios of the moments.

\section{Variance on small scales}
\label{sectvariance}

We now consider the effect of a (possibly) large contribution to the curvaton
density from smaller scale modes, compared with the cosmological
scales probed directly, for instance, by the CMB anisotropies. This
situation was recently discussed by Linde and Mukhanov \cite{LM06}
(see also \cite{LM97,Lyth06}). Such smaller scale modes might contribute
significantly to the average curvaton energy density on larger scales
if the curvaton field power spectrum rises on smaller scales, or if
some of the fraction (even if it is a small fraction) of the energy
from the inflaton decay at the end of inflation is transfered to the
curvaton \cite{LM06}. In either case we will describe this by a small
scale variance in the curvaton field up to some averaging scale
\begin{equation}
 \Delta_s^2 = \frac{\langle \delta\cur^2 \rangle_s}{\bar\cur^2} \,.
\end{equation}
The key observation is that these small-scale field fluctuations on
spatially-flat hypersurfaces are uncorrelated with the field
perturbations on larger scales.
Thus there is an additional contribution to the average curvaton
energy density
\begin{equation}
\bar\rho_\cur= \frac12 m^2 (1+\Delta_s^2) \bar\cur^2 \,,
\end{equation}
which is {\em homogeneous} on large
scales. In effect the curvaton density can be spilt into two parts:
one that is perturbed on large scales, and one that is not.

We can include the contribution from this small-scale variance in our
non-linear expression (\ref{nlzeta}) for the curvature perturbation,
$\zeta$, in the sudden-decay approximation where the
curvaton decays on a uniform-density hypersurface, to give the equation
\begin{eqnarray}
 \label{nlzetaDelta}
(1-\Omsd) e^{-4\zeta}
& + & \frac{1}{1+\Delta_s^2}\Omsd e^{3(\zeta_\cur-\zeta)}\nonumber\\
& + & \frac{\Delta_s^2}{1+\Delta_s^2} \Omsd e^{-3\zeta} = 1 \,,
\end{eqnarray}
where we have set to zero any pre-existing perturbation in the
radiation, $\zeta_r=0$.

At first order this shows how the resulting curvature perturbation on
large scales is suppressed by the small-scale variance:
\begin{equation}
\label{zeta1withvariance}
\zeta_1 = \frac{r}{1+\Delta_s^2} \zeta_{\cur1} \,,
\end{equation}
where $r$ is given by Eq.~(\ref{defrSD}) in the sudden-decay
approximation, and we use $\zeta_{\cur1}$ to denote the fractional
field perturbation on large scales, given in Eq.~(\ref{zetas1}).

However small-scale variance also affects the non-Gaussianity at second- and
higher-orders. At second-order we find
\begin{equation}
 \zeta_2 = \left[ \frac{3(1+\Delta_s^2)}{2r} \left( 1 +
 \frac{gg''}{g^{\prime2}} \right) -2 -r \right] \zeta_1^2 \,,
\end{equation}
and thus we have
\begin{equation}
 \fNL = (1+\Delta_s^2) \frac{5}{4r} \left( 1 +
 \frac{gg''}{g^{\prime2}} \right)
- \frac53 -\frac{5r}{6} \,.
\end{equation}
Note that the small-scale variance, although homogeneous,
is not equivalent to additional homogeneous radiation
due to its non-relativistic equation of state.

If we allow for non-instantaneous curvaton decay we find
\begin{equation}
\label{fnllargeD}
 \fNL = (1+\Delta_s^2) \frac{5}{4r} \left( 1 +
 \frac{gg''}{g^{\prime2}} \right)
+ \frac{5}{6} \frac{d^2N/dp^2}{\left( dN/dp\right)^2} \,,
\end{equation}
where any $\Delta_s^2$ dependence in the second term on the
right-hand-side cancels out
so that we can use the numerical results presented in 
Sec.~\ref{sectnumerical} to evaluate this term.

\subsection{Observational constraints on small-scale variance}

The fact that the non-linearity parameter grows with the small-scale
variance means that we can constrain the small-scale variance
from constraints on the non-linearity parameter $\fNL$ on larger
scales.

In practice the non-linearity at each successive order still depends
on the non-linear evolution function for the curvaton field, $g(\cur
_*)$. Hence we cannot rule out models where the small-scale variance
is large, but its effect is precisely cancelled by the non-linear
evolution. For simplicity we assume in the following that the
non-linear evolution is negligible so that $g''$ and higher
derivatives can be set to zero.

Recalling, that $r\le 1$ and $-54 < \fNL < 114$ (from WMAP3 \cite{Spergel:2006hy}) we
find an upper bound 
\begin{equation}
\Delta_s^2 < 90 \,.
\end{equation}

\subsection{Observational constraints on variance on CMB scales}

On the scales directly probed by CMB observations, the constraint on
the variance will be much tighter, since 
in addition to $\fNL$ constraint we observe
\begin{equation}
  \langle \zeta_1^2\rangle_{\rm CMB} = A^2\,,
\end{equation}
with $A^2 \simeq 6.25\times10^{-10}$. Substituting
(\ref{zeta1withvariance}) with (\ref{zetas1}) into the left hand side
we get
\begin{equation}
\label{DeltaCMBeqn}
\frac{4}{9} \frac{\Delta_{\rm CMB}^2}{(1+\Delta_s^2)^2} r^2 = A^2 \,,
\end{equation}
where
\begin{equation}
\Delta_{\rm CMB}^2 = \frac{\langle\delta\cur^2\rangle_{\rm CMB}}{\bar\cur^2}\,.
\end{equation}
Equation (\ref{DeltaCMBeqn}) gives
\begin{equation}
\Delta^2_{\rm CMB} = \frac{9}{4} A^2 \frac{(1+\Delta_s^2)^2}{r^2} \,,
\end{equation}
and eliminating $(1+\Delta_s^2)^2/r^2$ with help of (\ref{fnllargeD}) we end up with
\begin{equation}
\Delta^2_{\rm CMB} = \frac{9}{4} A^2 \left( \frac{4}{5}\fNL
-\frac{2}{3} \frac{d^2N/dp^2}{\left( dN/dp\right)^2}  \right)^2 \,. 
\end{equation}
The maximum of the absolute value of the second term in the parenthesis is
numerically found to be
always less than $2$. Thus, employing the triangle inequality, we find
\begin{equation}
\Delta^2_{\rm CMB} < \frac{9}{4} A^2 \left( |\frac{4}{5}\fNL| + 2
\right)^2 \,. 
\end{equation}
But the WMAP3 upper limit for $|\frac{4}{5}\fNL|$ is $91$, which implies
\begin{equation}
\Delta^2_{\rm CMB} < \frac{9}{4} A^2 \times 93^2 = 1.2 \times 10^{-5} \,.
\end{equation}

\section{Conclusions}
\label{sectconclusions}


In this paper we have presented for the first time the fully
non-linear probability density function (pdf) for the
primordial curvature perturbation on large scales in the curvaton
scenario using the $\delta N$-formalism.
By solving the non-linear evolution equations in an unperturbed
(FRW) universe one can construct the local expansion upto a final
uniform density as a function of the initial curvaton field value,
$N(\cur)$. Assuming a Gaussian form for the initial field
distribution on large scales (as would be expected for a weakly
coupled scalar field after inflation) it is straightforward to
construct the probability density function for $\delta N$ and
hence the non-linear curvature perturbation $\zeta$, defined in
Eq.~(\ref{zetanl}).
This procedure is particularly simple in the case where the local
expansion is a function of a single scalar field, such as the
curvaton, but it is also straightforward to apply to multiple fields
whose initial distributions on large scales are known.

In the sudden-decay approximation where it is assumed that the
curvaton decays instantaneously, when $H\sim\Gamma$, we have
presented a simple non-linear analytic expression,
Eq.~(\ref{nlzeta}), relating the primordial curvature perturbation
to the initial curvaton perturbation.
We have compared analytic results in the sudden-decay approximation,
with our results derived from direct numerical integration of the
full coupled equations for the local radiation and curvaton energy
densities and found good quantitative agreement.

In particular we have calculated the leading-order contributions to
the primordial bispectrum and trispectrum, including for the first
time the effect of third-order terms in the curvature perturbation.
In some cases \cite{LythNG,nurmi} non-linear evolution of the
curvaton field on super-Hubble scales, after Hubble-exit during
inflation, but before the curvaton begins to oscillate about the
minimum of its potential, could lead to a suppression of the leading
order contribution to the primordial bispectrum. We have shown that
in this case there will instead be a large contribution to the
primordial trispectrum, unless there is an additional cancellation
in the third-order term.

We have computed numerically the moments of the pdf for the primordial
curvature perturbation up to and including the sixth-order moment for
a range of values of the linear transfer coefficient, $r$. To
accurately reproduce the even moments of the distribution we need to
go beyond the second-order terms in the curvature perturbation
(described by the non-linearity parameter $\fNL$) and include
higher-order terms.

One example of how non-Gaussianity can be used to constrain model
parameters is the case when the curvaton field has a large variance
on small scales, as recently proposed by Linde and Mukhanov
\cite{LM06}. In this case the suppression of the linear curvature
perturbation is accompanied by an increase in non-Gaussianity. We
have shown that in this case limits on the primordial bispectrum can
be used to place limits on the small scale variance.

The calculations presented here should enable the curvaton model to
be subjected to a range of tests of non-Gaussianity, going beyond
just the bispectrum. In the simplest models (neglecting non-linear
evolution of the field before it decays) the non-Gaussianity is a
function of a single parameter, $r$, which is the linear transfer
coefficient relating the first-order primordial curvature
perturbation with the curvaton perturbation at Hubble-exit during
inflation. Multiple tests of the form of any primordial
non-Gaussianity could offer consistency tests of the curvaton
scenario.

\begin{acknowledgments}
  JV thanks Sami Nurmi and Bj\"orn Malte Sch\"afer for useful
  discussions and Ossi Pasanen for teaching basic Maple programming.
  DW is grateful to David Lyth and Karim Malik for useful discussions,
  and is grateful to the Yukawa Institute for Theoretical Physics,
  Kyoto University , for its hospitality when this work was begun
  during YKIS2005 and the post-YKIS workshop in July 2005.  JV and DW
  are supported by PPARC grant PP/C502514/1.  MS is supported by JSPS
  Grant-in-Aid for Scientific Research(S) No.~14102004 and (B)
  No.~17340075.

\end{acknowledgments}


\appendix
\section{}
\label{sectappendix}

In this appendix we solve the primordial curvature perturbation $\zeta$ as a
function of initial Gaussian field perturbation $\delta_1\cur$ in
the sudden-decay approximation. We also solve the inverse problem, i.e.,
find $\delta_1\cur/\bar\cur$ (or $\zeta_1$) as a function of $\zeta$. Using these
results we derive an analytic expression for the (non-Gaussian) probability
density function of $\zeta$; $\tilde f(\zeta)$.

We can rewrite Eq.~(\ref{nlzetasimpl}) in the form
\begin{equation}
\label{nlzetaeqnappendix}
e^{4\zeta} - \left[ \frac{4r}{3+r} e^{3\zeta_\cur} \right] e^{\zeta} + \left[ \frac{3r-3}{3+r}\right] = 0 \,,
\end{equation}
where $r = 3\Omsd/(4-\Omsd)$. This is a fourth degree equation for $X=e^\zeta$.
The solution of this full non-linear equation which gives the primordial curvature
perturbation as a function of initial Gaussian curvaton field $\cur_\ast({\bf x})$ is
\begin{equation}
\label{zetafullAppendix}
  \zeta = \ln(X)\,,
\end{equation}
with
\begin{equation}
X = K^{1/2}\frac{1+\sqrt{ArK^{-3/2}-1}}{(3+r)^{1/3}} \,,
\end{equation}
where

\vspace{-7mm}
\begin{eqnarray}
A & \definit & e^{3\zeta_\cur}
= \frac{\rho_{\cur, \rm osc}({\bf x})}{\bar\rho_{\cur, \rm osc}} = \left[ \frac{g(\cur_\ast({\bf x}))}{g(\bar\cur_\ast)} \right]^2 \,,\\
K & \definit & \half \left[ P^{1/3} +(r-1)(3+r)^{1/3} P^{-1/3} \right] \,,\\
P & \definit & (Ar)^2 + \left[ (Ar)^4 - (3+r)(r-1)^3 \right]^{1/2} \,.
\end{eqnarray}

The inverse problem (solving initial $\cur/\bar\cur$ as a function of $\zeta$) is
much simpler. Namely, Eq.~(\ref{nlzetaeqnappendix}) gives immediately
\begin{equation}
\label{Eqinverseproblem}
e^{3\zeta_\cur} = \frac{3+r}{4r} e^{3\zeta} + \frac{3r-3}{4r} e^{-\zeta} \,,
\end{equation}
and here $e^{3\zeta_\cur} = g^2[\cur_\ast({\bf x})]/g^2(\bar\cur_\ast) = \cur^2({\bf x})/\bar\cur^2$.

Assuming that there is {\it no non-linear evolution} between
the Hubble exit and start of curvaton oscillation [$g^{(n)} = 0$ for $n>1$]
the left hand side of (\ref{Eqinverseproblem}) is  {\it exactly} [see Eq.~(\ref{Eqfullzetacur})]
\begin{equation}
e^{3\zeta_\cur} = \left( 1 + \frac{\delta_1\cur}{\bar\cur}\right)^2
= 1 + 2 \frac{\delta_1\cur}{\bar\cur} + \left(\frac{\delta_1\cur}{\bar\cur}\right)^2 \,.
\end{equation}
Hence (\ref{Eqinverseproblem}) simplifies to a second degree equation for
$\delta_1\cur/\bar\cur$. The solutions are
\begin{equation}
\label{Eqdeltacurappendix}
\left(\frac{\delta_1\cur}{\bar\cur}\right)_{\pm} = -1 \pm \left[  \frac{3+r}{4r} e^{3\zeta} + \frac{3r-3}{4r} e^{-\zeta} \right]^{1/2} \,,
\end{equation}
where the ``$+$'' sign corresponds to a small perturbation and ``$-$'' sign
would give $|\delta_1\cur/\bar\cur| \sim 1$.
An alternative Gaussian ``reference variable'' is the linear end result $\zeta_1$.
From (\ref{defr}) and (\ref{zetas1}) we have $\zeta_1 = \frac{2}{3}r \frac{\delta_1\cur}{\bar\cur}$.
In Sec.~\ref{sectpdf} we will need a derivative of this Gaussian
random variable $\zeta_1$ with respect to $\zeta$.
Using (\ref{Eqdeltacurappendix}) we easily find
\begin{eqnarray}
\frac{d\zeta_{1\pm}}{d\zeta}
& = & \pm\frac{1}{3} r \left[  3\frac{3+r}{4r} e^{3\zeta} - \frac{3r-3}{4r} e^{-\zeta} \right]\nonumber\\
&& \ \ \ \ \times\left[  \frac{3+r}{4r} e^{3\zeta} + \frac{3r-3}{4r} e^{-\zeta} \right]^{-1/2}\!\!\!\!\!\!\!.
\end{eqnarray}
Hence the full non-Gaussian probability density function for $\zeta$ is
\begin{equation}
\label{fullsdpdf}
\tilde f(\zeta) = \tilde f_{-}(\zeta) + \tilde f_{+}(\zeta) \,,
\end{equation}
where
\begin{eqnarray}
\tilde f_{\pm}(\zeta) & = & \left|\frac{d\zeta_{1\pm}}{d\zeta}\right| f_g(\zeta_{1\pm}) \,,
\end{eqnarray}
and
\begin{equation}
f_g(\zeta_{1\pm}) =
\frac{1}{\sqrt{2\pi\sigma_g^2}}e^{-\zeta_{1\pm}^2/(2\sigma_g^2)} 
\end{equation}
with $\zeta_{1\pm}$ being $2r/3$ times the rhs of
Eq.~(\ref{Eqdeltacurappendix}). Here $f_g(\zeta_1)$ is the Gaussian
pdf for the first order perturbation $\zeta_1$ with variance
$\sigma_g^2 = \sigma_{\zeta_1}^2$ ($\sim 6.25\times10^{-10}$ to match
the observations), and mean $\mu = 0$. In practice, $ \tilde
f_{-}(\zeta)$ could be neglected, because $f_g(\zeta_{1-})$ is
typically of the order $\exp(-10^{10})$.  Substituting all
ingredients into (\ref{fullsdpdf}) the pdf reads
\begin{widetext}
\begin{eqnarray}
\tilde f_{\rm SD}(\zeta) & = & \frac{1}{\sqrt{2\pi\sigma_g^2}}
\frac{1}{2}\Big[ (3+r) e^{3\zeta} + (1-r) e^{-\zeta} \Big] \left[
\frac{3+r}{r} e^{3\zeta} + \frac{3r-3}{r} e^{-\zeta} \right]^{-1/2}
\nonumber\\ 
& \times & \sum_{\pm} \exp\left\{-\frac{4}{9}r^2 \left[-1
\pm \left( \frac{3+r}{4r} e^{3\zeta} + \frac{3r-3}{4r} e^{-\zeta}
\right)^{1/2}\right]^2 \Big/(2\sigma_g^2)\right\} \,,
\label{fullsdpdf2}
\end{eqnarray}
where the subscript SD reminds us that this is the sudden-decay result.
\end{widetext}

\bibliography{ngcurvatonarxiv3}

\end{document}